\documentclass[pra,preprint,superscriptaddress,amsmath,amssymb,showpacs]{revtex4}
\usepackage{bm}
\usepackage{amsmath}
\usepackage[dvips]{graphicx}
\begin{document}

\title{Photon splitting in a laser field}

\author{A. Di Piazza}
\email{dipiazza@mpi-hd.mpg.de}
\affiliation{Max-Planck-Institut f\"ur Kernphysik, Saupfercheckweg 1, 69117 Heidelberg, Germany}
\author{A. I. Milstein}
\email{milstein@inp.nsk.su}
\affiliation{Max-Planck-Institut f\"ur Kernphysik, Saupfercheckweg 1, 69117 Heidelberg, Germany}
\affiliation{Budker Institute of Nuclear Physics, 630090 Novosibirsk, Russia}
\author{C. H. Keitel}
\email{keitel@mpi-hd.mpg.de}
\affiliation{Max-Planck-Institut f\"ur Kernphysik, Saupfercheckweg 1, 69117 Heidelberg, Germany}

\date{\today}

\begin{abstract}
Photon splitting due to vacuum polarization in a laser field is considered.
Using  an operator technique, we derive the amplitudes for arbitrary strength, spectral content and polarization of  
the laser field. The case of a monochromatic circularly polarized laser field is studied in detail and the amplitudes are obtained as three-fold integrals. The asymptotic behavior of the amplitudes for various limits of interest are investigated also in the case of a linearly polarized laser field. Using the obtained results, the possibility of experimental observation of the process is discussed.
\end{abstract}
 
\pacs{12.20.Ds (QED Specific Calculations), 42.62.-b (Laser Applications), 42.50.Xa (Optical Tests in Quantum Electrodynamics)} 
\maketitle
\section{Introduction}
Vacuum polarization is known to result in such nonlinear Quantum Electrodynamics (QED) effects as coherent photon scattering and  photon splitting in an external electromagnetic field. Photon scattering in the electric field of atoms (Delbr\"uck scattering) has been investigated in detail both theoretically and experimentally \cite{PM75,MS94,GS95,Delbexp}. Recently, an essential progress in the understanding of the photon splitting process
in an atomic electric field was achieved \cite{LMS,LMMST03} and the first successful observation of photon splitting was performed \cite{splitexp}. It turned out that higher orders of the perturbation theory with respect to $Z\alpha$ play an important role and drastically modify the cross sections of Delbr\"uck scattering and photon splitting at high photon energy (here, Z is the nuclear charge number, $\alpha=e^2=1/137$ is the fine-structure constant with $e$ being the electron charge, the system of units $\hbar=c=1$ is used). Therefore, the theoretical and experimental investigation of photon splitting gives the possibility to verify QED in the presence of strong external field when the nonperturbative effects in the field strength (contributions of higher-order terms) are very important.

Photon splitting in a constant and uniform electromagnetic field for arbitrary values of the field strength was considered in numerous papers because of the potential astrophysical implications of the process (see, e.g., Refs. \cite{Adler70,BB70,Adler71,Papanyan74,BMS869697}). The possibility of observing photon splitting in the electric fields of single crystals at high photon energies was investigated in \cite{BMS87}. Photon splitting due to vacuum polarization in a strongly magnetized plasma was discussed in \cite{Marklund}.

Photon splitting in a laser field with arbitrary strength and frequency content has never been considered before for an arbitrary energy of the incoming photon. The kinematics of photon splitting depends essentially on the structure of the external field. In the Coulomb field, the wave vector $\bm k_1$ of the initial photon and the wave vectors $\bm k_2$ and  $\bm k_3$ of the final ones are, in general, not coplanar. In a constant and uniform electromagnetic field, the initial and final photons are almost collinear (up to nonzero photon mass due to vacuum polarization in the external field). In a classical  electromagnetic plane wave, due to energy-momentum conservation, the initial and final photons are 
coplanar in the frame where the laser wave propagates in the direction anti-parallel to $\bm k_1$. Throughout the paper we refer to this frame as the laboratory frame. The structure of the photon splitting amplitude in a laser field depends on the value of two Lorentz-invariant parameters. In the laboratory frame they are given by
\begin{equation}\label{chieta}
\eta=\frac{\omega_1\omega_0}{m^2}\, ,\quad \chi=\frac{\omega_1}{m}\frac{E}{E_c}\, ,
\end{equation}
where $\omega_1$ is the energy of the initial photon, $\omega_0$ and $E$ are the characteristic frequency and electric field strength of the laser pulse, $E_c=m^2/|e|=1.3\times 10^{16}\,$V/cm is the critical electric field and $m$ is the electron  mass. For a monochromatic laser field, $\omega_0$ and $E$ in Eq. (\ref{chieta}) are the laser frequency and the root mean square of the laser electric field. For $\eta\ll 1$ and $\chi\ll 1$, photon splitting in a monochromatic laser wave was considered in \cite{Aff87} in the leading approximation with the help of the Euler-Heisenberg Lagrangian. The contribution to the amplitudes of photon splitting linear in $\chi$ but at arbitrary $\eta$ is nothing but the usual photon-photon scattering which was studied in detail theoretically (see, e.g., \cite{LL}) but never observed experimentally \cite{Moulin}. The possibility to observe ``laser-assisted'' photon-photon scattering by using the next generation of petawatt lasers has been investigated in \cite{Di_Piazza_2005,Lundstroem_2006}. Other processes related to photon-photon scattering and initiated by vacuum polarization in the presence of the superposition of two laser fields, or a laser field and a Coulomb field have been considered in \cite{Aleksandrov_1986,Ivan,Di_Piazza_2006,Heinzl_2006}.

For the strongest optical ($\omega_0\approx 1\,$eV) laser beam available today the laser intensity is $I\approx 10^{22}\,\text{W/cm}^2$ and the corresponding electric field amplitude is $E\approx 10^{-4}E_c$ \cite{Emax}. The Extreme Light Infrastructure (ELI) project aims at reaching laser intensity values higher than $10^{25}\,\mbox{W/cm}^2$ corresponding to $E\approx 10^{-2}E_c$. The X-ray Free-Electron Laser (X-FEL) facilities presently being developed at DESY (Hamburg, Germany) and at SLAC (Stanford, USA) are intended to provide 
spatially coherent and highly brilliant beams of synchrotron radiation
with single-photon energies up to $8\div 12$ keV at a maximum intensity $\approx 10^{15}\mbox{W/cm}^2$ corresponding to a field strength of 
$E\approx 10^{-7}E_c$ \cite{XFEL}. The above mentioned projects have initiated intensive investigations of various QED processes that can be experimentally observed when using these future facilities \cite{ChristophReview,MarklundReview,MourouReview}. On the other hand, sources of high-energy photons are nowadays available in a wide range of photon energies spanning from tens of MeV up to hundreds of GeV. It will then be possible to reach experimentally unexplored regions for the values of the parameters $\eta$ and $\chi$ where vacuum polarization effects in a laser field, and photon splitting in particular, are essentially nonperturbative.

In the present paper, we derive the amplitudes of photon splitting in a plane wave described by an arbitrary four-vector potential $A^{\mu}(\phi)=(0,\bm A(\phi))$, where $\phi=\varkappa x$ with $\varkappa=(1,\bm\varkappa)$, $\varkappa^2=0$, and $\bm\varkappa\cdot\bm A(\phi)=0$. At a first stage, we do not assume that the plane wave is monochromatic or even
periodic. It is convenient, without loss of generality, to consider the process in the laboratory frame, i.e. in the frame where $\bm k_1$ is anti-parallel to the vector $\bm\varkappa$. The calculations are performed by using the operator technique developed in Refs. \cite{BMS75, BKMS75} for the calculation of the polarization and mass operators in a laser field. Alternative forms of these operators were derived independently in Refs. \cite{BM75,BM76} by means of a different method. The results of Refs. \cite{BMS75, BKMS75} and Refs. \cite{BM75,BM76} are in agreement with each other. The operator technique allows one to find the amplitudes without using the explicit form of the Green's function for the Dirac equation in the external field. This circumstance considerably simplifies the calculations.

The rest of the paper is organized as follows: in Section II we describe the general properties of the photon splitting amplitudes. Then, starting from the operator representation of the amplitudes, in Sections III and IV we calculate the explicit expressions of the amplitudes for a general vector potential of the form $\bm A(\phi)$. In Section V we consider a monochromatic laser field with particular emphasis on the case of a circularly polarized laser field. In Section VI we analyze various asymptotic forms of the amplitudes both for a circular and for a linearly polarized field. The possibility of the experimental observation of photon splitting in a laser field is discussed in detail in Section VII. Finally, in Section VIII the main conclusions of the paper are presented.

\section{General discussion}
\begin{figure}[ht]
\begin{center}
\includegraphics[width=8cm]{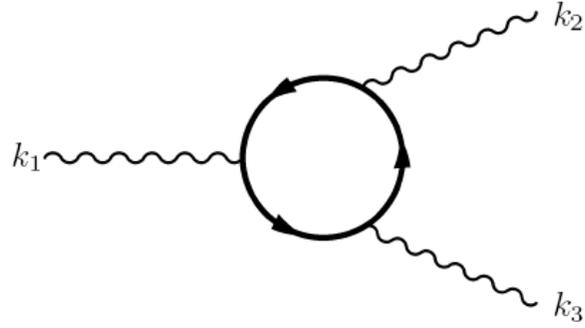}
\caption{\label{PSD} Feynman diagram for the photon splitting amplitude given by Eq. (\ref{MG}). This diagram corresponds to the Furry representation and the thick line denotes the electron propagator (Green's function) in the laser field. The wavy lines symbolize the external photons. The diagram with the permutation $k_2\leftrightarrow k_3 $ has to be added.} 
\end{center} 
\end{figure}

The amplitude $M$ of the photon splitting process of a photon with four-momentum $k_1=(\omega_1,{\bm k}_1)$ into two photons with four-momenta $k_2=(\omega_2,{\bm k}_2)$ and $k_3=(\omega_3,{\bm k}_3)$ with $k_1^2=k_2^2=k_3^2=0$ is represented by the Feynman diagram shown in Fig. \ref{PSD}. The corresponding formal expression for $M$ in the Furry representation reads
\begin{eqnarray}\label{MG}
M&=&ie^3\int\,d^4 x\,\mbox{Tr}\langle x|
\mbox{e}^{-ik_1x}\,\hat e_1\frac{1}{\hat{\mathcal{P}}-m+i0}\mbox{e}^{ik_2x}\,\hat e_2^*\frac{1}{\hat{\mathcal{P}}-m+i0}
\mbox{e}^{ik_3x}\,\hat e_3^*\frac{1}{\hat{\mathcal{P}}-m+i0}|x\rangle\,\nonumber\\
&&+ (k_2\leftrightarrow k_3\, , \,e_2\leftrightarrow e_3)\, ,
\end{eqnarray}
where $e_{1\mu}$ is the initial photon polarization four-vector, $e_{2\mu}$ and $e_{3\mu}$ are the polarization four-vectors of the final photons, $\hat {\mathcal P}={\mathcal P}_\mu\gamma^\mu$ with $\mathcal{P}_\mu=i\partial_\mu-eA_\mu(x)$ and $\gamma^\mu$ being the Dirac matrices. The factor $(4\pi)^{3/2}$, contained in the external photon wave functions, has been included into the formula for the probability of the process. The calculation of the amplitude $M$ is essentially simplified by using the Green's function $D(x_2,\, x_1)$ of the ``squared'' Dirac equation:
\begin{equation}
D(x_2,\,x_1)=\langle
x_2|\frac{1}{\hat{\mathcal{P}}^2-m^2+i0}|x_1\rangle\, .
\end{equation}
As shown in \cite{LMS}, the expression of $M$ can be written as a sum of two contributions containing either three or two Green's functions $D(x_2,\,x_1)$:
\begin{equation}\label{tensorM}
M\equiv{\cal M}^{\mu\nu\rho}e_{1\mu}e_{2\nu}^*e_{3\rho}^*=M^{(2)}+M^{(3)}\,.
\end{equation}
The term $M^{(2)}$ has the form
\begin{eqnarray}\label{M2}
&&M^{(2)}=M_{1}^{(2)}+M_{2}^{(2)}+M_{3}^{(2)}\\
&&M_{1}^{(2)}=-ie^3(e_2^*e_3^*)\!\!\int\!\! d^4 x\mbox{Tr}\langle x|
\mbox{e}^{-ik_1x}(2e_1{\mathcal
P}-\hat e_1\hat k_1)\frac{1}{\hat{\mathcal{P}}^2-m^2+i0}
\mbox{e}^{i(k_2+k_3)x}\frac{1}{\hat{\mathcal{P}}^2-m^2+i0}
|x\rangle\,,\nonumber\\
&&M_{2}^{(2)}=-ie^3(e_1e_3^*)\!\!\int\!\! d^4 x\mbox{Tr}\langle x|
\mbox{e}^{ik_2x}(2e_2^*{\mathcal
P}+\hat e_2^*\hat k_2)\frac{1}{\hat{\mathcal{P}}^2-m^2+i0}
\mbox{e}^{i(k_3-k_1)x}\frac{1}{\hat{\mathcal{P}}^2-m^2+i0}
|x\rangle\,,\nonumber\\
&& M_{3}^{(2)}=-ie^3(e_1e_2^*)\!\!\int\!\! d^4 x\mbox{Tr}\langle x|
\mbox{e}^{ik_3x}(2e_3^*{\mathcal
P}+\hat e_3^*\hat k_3)\frac{1}{\hat{\mathcal{P}}^2-m^2+i0}
\mbox{e}^{i(k_2-k_1)x}\frac{1}{\hat{\mathcal{P}}^2-m^2+i0}
|x\rangle\,.\nonumber
\end{eqnarray}
The term $M^{(3)}$ is given by
\begin{eqnarray}\label{M3}
M^{(3)}&=&\frac{i}{2}{e^3}\int\,dx\,\mbox{Tr}\langle x|
\mbox{e}^{-ik_1x}(-\hat e_1\hat
k_1+2e_1{\mathcal P})\frac{1}{\hat{\mathcal{P}}^2-m^2+i0}\nonumber\\
&&\times\mbox{e}^{ik_2x}(\hat e_2^*\hat
k_2+2e_2^*\mathcal{P})\frac{1}{\hat{\mathcal{P}}^2-m^2+i0}
\mbox{e}^{ik_3x}(\hat e_3^*\hat
k_3+2e_3^*\mathcal{P})\frac{1}{\hat{\mathcal{P}}^2-m^2+i0}|x\rangle\,\nonumber\\
&&+ (k_2\leftrightarrow k_3\, , \,e_2\leftrightarrow e_3)\, .
\end{eqnarray}

For each photon with the four-momentum  $k^\mu_i$ ($i=1,2,3$) it is convenient
to introduce two vectors $\varepsilon_{i,\lambda_i}^\mu$, ($\lambda_i=1,2$)
\begin{eqnarray}\label{varepsilon}
&&\varepsilon_{i,\lambda_i}^\mu=a_{\lambda_i}^\mu-\frac{k_ia_{\lambda_i}}{k_i\varkappa}
\varkappa^\mu\, ,\quad  \varepsilon_{i,\lambda_i}\varepsilon_{j,\lambda_j}= -\delta_{\lambda_i\lambda_j}\,\nonumber\\
&&a_\lambda^0=0\,,\quad ({\bm a}_\lambda)^2=1\, ,\quad {\bm a}_1\cdot{\bm
a}_2=0\,,\quad {\bm a}_\lambda\cdot {\bm\varkappa}=0.
\end{eqnarray}
Due to gauge invariance, it is possible to write the tensor ${\cal M}^{\mu\nu\rho}$ introduced in Eq. (\ref{tensorM}) for real photons
 in the form (see Ref. \cite{BMS75}):
\begin{equation}
{\cal M}^{\mu\nu\rho}=-\sum_{\lambda_1,\lambda_2\lambda_3=1,2}
R_{\lambda_1\lambda_2\lambda_3}\varepsilon_{1,\lambda_1}^\mu
\varepsilon_{2,\lambda_2}^\nu\varepsilon_{3,\lambda_3}^\rho \, ,
\end{equation}
For the coefficients $R_{\lambda_1\lambda_2\lambda_3}$ we have
\begin{equation}\label{RVM}
R_{\lambda_1\lambda_2\lambda_3}={\cal M}_{\mu\nu\rho}
\varepsilon_{1,\lambda_1}^\mu
\varepsilon_{2,\lambda_2}^\nu\varepsilon_{3,\lambda_3}^\rho \, .
\end{equation}
So, the coefficients $R_{\lambda_1\lambda_2\lambda_3}$ can be
calculated from Eqs. (\ref{M2}) and (\ref{M3}) by the substitution
$e_i^\mu\longrightarrow\varepsilon_{i,\lambda_i}^\mu$.
Using the  coefficients $R_{\lambda_1\lambda_2\lambda_3}$, we write the amplitude $M$ of photon splitting in the laser field as
\begin{equation}\label{MviaS}
M=-\sum_{\lambda_1,\lambda_2\lambda_3=1,2}
R_{\lambda_1\lambda_2\lambda_3}(\varepsilon_{1,\lambda_1}e_1)
(\varepsilon_{2,\lambda_2}e_2^*)(\varepsilon_{3,\lambda_3}e_3^*) \, .
\end{equation}
This expression of the amplitude $M$ is very convenient because it automatically provides the fulfillment of the gauge invariance principle. As a consequence, it is not necessary to perform the subtraction in the calculation of the lowest-order photon splitting amplitude, described by the photon-photon scattering diagrams.

We perform the calculations in terms of the helicity amplitudes $M_{\sigma_1\sigma_2\sigma_3}$, with $\sigma_j=\pm 1$. In this case, the polarization vector $\bm e_{\sigma}$ of each external photon with the wave vector $\mathbf{k}$ and the energy $\omega$ satisfies the relations $\bm e_\sigma\cdot\bm k=0$ and $\bm e_{\sigma}\times\bm k=i\sigma\omega \bm e_\sigma$. Due to momentum conservation in the laser field, the final transverse momenta $\bm k_{2\perp}$ and $\bm k_{3\perp}$ (with respect to the laser propagation direction) are equal in modulus and opposite in direction:
$\bm k_{2\perp} = -\bm k_{3\perp}\equiv\bm q$ and $\bm q\cdot\bm\varkappa=\bm q\cdot\bm k_1=0$. We direct the vector $\bm a_1$ along the vector $\bm q$, so that $\bm a_2=\bm\varkappa\times\bm a_1$, then
\begin{eqnarray}
\bm e_{1\,,\sigma_1}&=&-\frac{1}{\sqrt{2}}(\bm a_2+i\sigma_1\bm a_1)\, ,\nonumber\\
\bm e_{2\,,\sigma_2}&=&\frac{1}{\sqrt{2}}(\bm a_2-i\sigma_2 \bm a_2\times\bm k_2/\omega_2)\, ,\nonumber\\
\bm e_{3\,,\sigma_3}&=&\frac{1}{\sqrt{2}}(\bm a_2-i\sigma_3 \bm a_2\times\bm k_3/\omega_3)\, .
\end{eqnarray}
The corresponding polarization four-vectors $e_{i,\sigma_i}=(0,\bm e_{i\,,\sigma_i})$ have the following products with the four-vectors $\varepsilon_{i,\lambda_i}$:
\begin{eqnarray}
\varepsilon_{1,\lambda_1}e_{1,\sigma_1}=\frac{1}{\sqrt{2}}(i\sigma_1\delta_{\lambda_1,1}+\delta_{\lambda_1,2})\, ,\nonumber\\
\varepsilon_{2,\lambda_2}e_{2,\sigma_2}=\frac{1}{\sqrt{2}}(i\sigma_2\delta_{\lambda_2,1}-\delta_{\lambda_2,2})\, ,\nonumber\\
\varepsilon_{3,\lambda_3}e_{3,\sigma_3}=\frac{1}{\sqrt{2}}(i\sigma_3\delta_{\lambda_3,1}-\delta_{\lambda_3,2})\, .
\end{eqnarray}
These useful relations allow us to state that the amplitude $M_{\sigma_1\sigma_2\sigma_3}$ can be obtained from the coefficient $R_{\lambda_1\lambda_2\lambda_3}$ by the substitution in the latter:
\begin{equation}\label{subs}
\bm a_{\lambda_1}\rightarrow -\frac{1}{\sqrt{2}}(i\sigma_1\bm a_1+\bm a_2)\, ,\quad
\bm a_{\lambda_2}\rightarrow \frac{1}{\sqrt{2}}(-i\sigma_2\bm a_1+\bm a_2)\, ,\quad
\bm a_{\lambda_3}\rightarrow \frac{1}{\sqrt{2}}(-i\sigma_3\bm a_1+\bm a_2)\, .
\end{equation}

The energies $\omega_{2,3}$ and the parallel components $k_{2,3\,\parallel}=\bm\varkappa\cdot\bm k_{2,3}$ of the momenta of the final photons are functions of the transverse momentum $\bm q$ and of the energy $\Omega$ absorbed from the laser field, $\Omega=\omega_2+ \omega_3- \omega_1$. It is convenient to introduce the quantities 
\begin{equation} \label{nu}
\nu_{2,3}=\frac{\varkappa k_{2,3}}{\varkappa k_1}=\frac{1\mp\cos\theta_0}{2}\, , \quad
\cos\theta_0=\sqrt{1-\frac{\bm q^2}{\omega_1\Omega}}\, .
\end{equation}
The angle  $\theta_0$ coincides with the angle between the vectors $\bm k_2$ and $\bm\varkappa$ in the frame where  $\bm k_2=-\bm k_3$, 
so that $\bm q^2\le \omega_1\Omega $. Then
\begin{eqnarray}\label{theta0}
&&\omega_2=\omega_1\nu_2+\Omega\nu_3\, ,\quad
k_{2\parallel}=-\omega_1\nu_2+\Omega\nu_3\,\nonumber\\
&&\omega_3=\omega_1\nu_3+\Omega\nu_2\,,\quad
k_{3\parallel}=-\omega_1\nu_3+\Omega\nu_2\,.
\end{eqnarray}

It follows from  Eqs. (\ref{M2}) and (\ref{M3}) that, like the amplitude $M$, the coefficients $R_{\lambda_1\lambda_2\lambda_3}$ can also be presented as the sum
$R_{\lambda_1\lambda_2\lambda_3}=R_{\lambda_1\lambda_2\lambda_3}^{(2)}+R_{\lambda_1\lambda_2\lambda_3}^{(3)}$.
In order to explain the application of the operator technique to the calculation of the photon splitting amplitude, we present the derivation of the contribution $R_{\lambda_1\lambda_2\lambda_3}^{(2)}$. The calculation of $R_{\lambda_1\lambda_2\lambda_3}^{(3)}$ can be performed in a similar way and only the final result is given for this contribution.

\section{Calculation of $R_{\lambda_1\lambda_2\lambda_3}^{(2)}$}

We direct the vector $\bm\varkappa$ along the z-axis and pass from the variables $t$ and $z$ to the variables
$\phi=\varkappa x=t-z$ and $ T=(t+z)/2$. In this way
$$p^0=i\partial_t=-p_\phi-p_T/2\, ,\quad
p^z=-i\partial_z=-p_\phi+p_T/2\,,\quad
p_\phi=-i\partial_\phi\,,\quad p_T=-i\partial_T.$$
Then, we use in Eq. (\ref{M2}) the representation of the operator Green's function derived in \cite{BKMS75}
\begin{eqnarray}\label{GR}
&&\frac{1}{\hat{\mathcal{P}}^2-m^2+i0}=-i\int_0^\infty
ds\,\exp[is(\hat{\mathcal{P}}^2-m^2)]\nonumber\\
&&=-i\int_0^\infty
ds\,\exp(-ism^2)\,\left\{1+\frac{e\hat{\varkappa}}{2p_T}
[\hat{A}(\phi+2sp_T)-\hat{A}(\phi)]\right\}\nonumber\\
&&\times\exp\left[-i\int_0^sds'\left({\bm p}_\perp-e{\bm
A}(\phi+2s'p_T)\right)^2\right]\,\exp(2ip_\phi p_T)\, .
\end{eqnarray}
As before, the index $\perp$ indicates transverse quantities with respect to the laser propagation direction $\bm\varkappa$. Now, we pass in Eq. (\ref{M2}) from the integration over the variables $T$ and $\bm x_{\perp}$ to the corresponding momenta $p_T$ and 
$\bm p_{\perp}$ in accordance with
$$\int dT\langle T|...|T\rangle\, \rightarrow \int\dfrac{dp_T}{2\pi}\langle p_T|...|p_T\rangle\,,\,
\int d\bm x_\perp \langle\bm x_\perp|...|\bm x_\perp\rangle\, \rightarrow \int\dfrac{d\bm p_\perp}{(2\pi)^2}\langle\bm p_\perp|...|\bm p_\perp\rangle\,.$$
We also exploit the relations
\begin{equation}
\exp(2isp_\phi p_T)g(\phi)\exp(-2isp_\phi p_T)=g(\phi+2sp_T)\,,\quad \exp(2isp_\phi p_T)|\phi\rangle =|\phi-2sp_T\rangle\, ,
\end{equation}
valid for an arbitrary function $g(\phi)$ of the variable $\phi$. Besides, the matrix elements $\langle p_T|...|p_T\rangle$ and $\langle\bm p_\perp|...|\bm p_\perp\rangle$ become
\begin{eqnarray}\label{delta}
\langle p_T|\exp[i2\omega_1T(\nu_2+\nu_3-1)|p_T\rangle =2\pi\delta\big(2\omega_1(\nu_2+\nu_3-1)\big)\, ,\nonumber\\
\langle\bm p_\perp|\exp[-i\bm x_\perp\cdot(\bm k_{2\perp}+\bm k_{3\perp})]|\bm p_\perp\rangle =(2\pi)^2\delta(\bm k_{2\perp}+\bm k_{3\perp})\,.
\end{eqnarray}
The factors (\ref{delta}) are common to all amplitudes. They will be omitted below in the expressions of the amplitudes and will be taken into account
 in the usual way in the formula for the probability of the process. The matrix element $\langle\phi|...|\phi\rangle$ also becomes the
$\delta$-function containing the variable $p_T$ in its argument. Therefore, the integration over $p_T$ becomes trivial as well as the integration over $\bm p_\perp$. Following the procedure pointed out above and using the useful relations
\begin{eqnarray}
&&k_2^{\mu}=\Omega\nu_3\varkappa^{\mu}+\nu_2k_1^{\mu}+q^{\mu}\, ,\quad
k_3^{\mu}=\Omega\nu_2\varkappa^{\mu}+\nu_3k_1^{\mu}-q^{\mu}\,,\nonumber\\
&&k_2k_3=2\Omega\omega_1\,,\quad k_2k_1=2\Omega\omega_1\nu_3\,,\quad k_3k_1=2\Omega\omega_1\nu_2\,,\quad
{\bm q}^2=4\Omega\omega_1\nu_2\nu_3\,,
\end{eqnarray}
with $q^{\mu}=(0,\mathbf{q})$, we arrive at the following result for $R_{\lambda_1\lambda_2\lambda_3}^{(2)}$:
\begin{eqnarray}\label{S2final}
&&R_{\lambda_1\lambda_2\lambda_3}^{(2)}=
-\frac{me^3}{(2\pi)^2}\sum_{j=1,2,3}\delta_{(j)}\iint\limits_0^{\quad\infty}\frac{ds_1\,ds_2}{(s_1+s_2)^2}
\exp[-i(s_1+s_2)]\nonumber\\
&&\times\int d\phi\exp\left\{i\Omega\phi
+i\frac{4\omega_1\Omega}{m^2}\varphi_j-i(s_1+s_2)\left[\int_0^1dy{\bm\Delta}^2(y\varphi_j)-
\left(\int_0^1dy{\bm\Delta}(y\varphi_j)\right)^2\right]\right\}\nonumber\\
&&\times\left[-2{\bm a}_{\lambda_j}\cdot\int_0^1dy
\bm{\Delta}(y\varphi_j)+\frac{(s_1+s_2)^2}{2s_1s_2}{\bm
a}_{\lambda_j}\cdot\bm{\Delta}(\varphi_j)\right]\, .
\end{eqnarray}
Here we defined the quantities
\begin{eqnarray}\label{S2notation}
&&\delta_{(1)}=\delta_{\lambda_2\lambda_3}\, ,\quad \delta_{(2)}=\delta_{\lambda_1\lambda_3}\, ,\quad
\delta_{(3)}=\delta_{\lambda_1\lambda_2}\, ,\nonumber\\
&&(\varphi_1,\,\varphi_{2},\,\varphi_{3})=\frac{s_1s_2}{s_1+s_2}(1,\,-\nu_2,\, -\nu_3)\,,
\quad {\bm\Delta}(u)=\frac{e}{m}[{\bm A}(\phi+4u\omega_1/m^2 )- {\bm A}(\phi)]\,. \quad
\end{eqnarray}
In the parametrization of the Green's function we  made the substitution $s\to s/m^2$. As it has to be according to the Furry theorem, the contribution (\ref{S2final}) contains only odd powers of the external field. The explicit result of the integration over $\phi$ depends on the form of the function $\bm A(\phi)$. For a monochromatic wave, the integral over $\phi$ can be easily performed (see Section V). 

\section{Calculation of $R_{\lambda_1\lambda_2\lambda_3}^{(3)}$}
The calculation of the contribution  $R_{\lambda_1\lambda_2\lambda_3}^{(3)}$ is similar to that of $R_{\lambda_1\lambda_2\lambda_3}^{(2)}$, but considerably more cumbersome. In order to bring the answer into a relatively short form, we introduce the convenient notation
\begin{eqnarray}
&&\quad S=s_1+s_2+s_3\, ,\,\, \tau_1= \frac{s_2\nu_2-s_3\nu_3}{ S}\,,\,\,
\tau_2=\nu_2-\tau_1\,,\quad
\tau_3=\nu_3+\tau_1\,,\,
\nonumber\\
&& \bm{ D}_1(y)= {\bm \Delta}(s_2\tau_2y+s_3\tau_3(1-y))\,, \,
\bm{ D}_2(y)= {\bm \Delta}(s_2\tau_2y)\,, \,
\bm{ D}_3(y)= {\bm \Delta}(s_3\tau_3y)\,,\nonumber\\
&&{\bm F}=\frac{1}{ S}\int_0^1dy[s_1\bm{ D}_1(y)+s_2\bm{
D}_2(y)+s_3\bm{ D}_3(y)]\,
,\nonumber\\
&&F_0=  S{\bm F}^2-\int_0^1dy[s_1\bm{ D}_1^2(y)+s_2\bm{
D}_2^2(y)+s_3\bm{ D}_3^2(y)]\, ,\nonumber\\
&& \bm G=\frac{2s_1}{ S}\int_0^1dy\,[s_2(\bm{ D}_1(y)-\bm{
D}_2(y))+s_3(\bm{ D}_1(y)-\bm{ D}_3(y))]\, ,
\end{eqnarray}
and use  the relation $s_1\tau_1=s_2\tau_2-s_3\tau_3$. Before taking the trace over the $\gamma$-matrices, we obtain the following compact representation for $R_{\lambda_1\lambda_2\lambda_3}^{(3)}$:
\begin{eqnarray}\label{S3trace}
&&R_{\lambda_1\lambda_2\lambda_3}^{(3)}=\frac{im e^3}{8\pi^2}\int\negthickspace\negthickspace\int\limits_0^{\infty}
\negthickspace\negthickspace\int\frac{ds_1\,ds_2\,ds_3}{ S^2}\int d\phi
\exp\left[i\left(\Omega\phi+\frac{4\omega_1\Omega s_2s_3}{m^2 S}+F_0- S +\bm \rho\cdot\bm G
\right)\right]\nonumber\\
&&\times\Bigg\{\frac{1}{4}{\mbox{Tr}}\Bigg[\left(-2{\bm a}_{\lambda_3}\cdot{\bm N}_3
+\frac{\hat{\varepsilon}_{3,\lambda_3}\hat{k}_3}{m}\right)\left(1+\frac{m\hat{\varkappa}\hat{d}_3}{2\omega_1}\right)
\left(-2{\bm a}_{\lambda_1}\cdot{\bm N}_1
-\frac{\hat{\varepsilon}_{1,\lambda_1}\hat{k}_1}{m}\right)\nonumber\\
&&\times\left(1+\frac{m\hat{\varkappa}\hat{
d}_2}{2\omega_1}\right)\left(-2{\bm a}_{\lambda_2}\cdot{\bm N}_2+\frac{\hat{\varepsilon}_{2,\lambda_2}\hat{k}_2}{m}\right)
\left(1+\frac{m\hat{\varkappa}\hat{d}_1}{2\omega_1}\right)\Bigg]\nonumber\\
&&+\frac{4i}{ S}\left[\delta_{(1)}{\bm a}_{\lambda_1}\cdot({\bm N}_1-\bm V)+
\delta_{(2)}{\bm a}_{\lambda_2}\cdot({\bm N}_2+\nu_2\bm V)+
\delta_{(3)}{\bm a}_{\lambda_3}\cdot({\bm N}_3+\nu_3\bm V)\right]\Bigg\}\, \nonumber\\
&&+ (2\leftrightarrow 3\, ,\,\, \bm \rho\rightarrow -\bm \rho )\,,
\end{eqnarray}
where
\begin{eqnarray}
&&\bm \rho=\frac{\bm q}{m}\, ,\quad {\bm N}_1={\bm F}-\frac{s_1\bm \rho}{ S}\,,\quad
{\bm N}_2=\bm F-\bm{
D}_2(1)+\frac{s_3\bm \rho}{ S \nu_2}\,,\nonumber\\
&&{\bm N}_3={\bm
F}-\bm{ D}_3(1)+\frac{s_2\bm \rho}{ S \nu_3}
\,,\quad \bm V=\frac{1}{2}(\bm d_1+\bm d_2+\bm d_3)\,,\nonumber\\
&&\bm d_1=\frac{\bm D_2(1)-\bm D_3(1)}{2\tau_1}\, ,\quad
\bm d_2=\frac{\bm D_2(1)}{2\tau_2}\, ,\quad
\bm d_3=\frac{\bm D_3(1)}{2\tau_3}\,.
\end{eqnarray}
The substitution $(2\leftrightarrow 3\, ,\,\, \bm \rho\rightarrow -\bm \rho )$ in Eq. (\ref{S3trace}) results in the replacement
${\bm N}_1(\bm \rho)\rightarrow {\bm N}_1(-\bm \rho)$, ${\bm N}_2(\bm \rho)\rightarrow {\bm N}_3(-\bm \rho)$, and
${\bm N}_3(\bm \rho)\rightarrow {\bm N}_2(-\bm \rho)$. In fact, the role of the term with the substitution $(2\leftrightarrow 3\, ,\,\, \bm \rho\rightarrow -\bm \rho ) $ is to provide the fulfillment of the Furry theorem ($R_{\lambda_1\lambda_2\lambda_3}^{(3)}$ contains only odd powers of the external field). Therefore, the sum of the two terms is twice the odd part of the first term with respect to the external vector potential ${\bm A}(\phi)$. Finally, we take  the trace and obtain:
\begin{eqnarray}\label{S3final}
&&R_{\lambda_1\lambda_2\lambda_3}^{(3)}=\frac{ime^3}{4\pi^2}\int\negthickspace\negthickspace\int\limits_0^{\infty}
\negthickspace\negthickspace\int\frac{ds_1\,ds_2\,ds_3}{ S^2}\int d\phi
\exp\left[i\left(\Omega\phi+\frac{4\omega_1\Omega s_2s_3}{m^2 S}+F_0- S +{\bm \rho}\cdot{\bm G}
\right)\right]\nonumber\\
&&\times\Bigg<-4({\bm a}_{\lambda_1}\cdot{\bm N}_1) ({\bm a}_{\lambda_2}\cdot{\bm N}_2)
({\bm a}_{\lambda_3}\cdot{\bm N}_3)+4({\bm a}_{\lambda_2}\cdot{\bm N}_2)
 ({\bm a}_{\lambda_3}\cdot{\bm N}_3)({\bm a}_{\lambda_1}\cdot{\bm V})
\nonumber\\
&&-4\nu_3({\bm a}_{\lambda_1}\cdot{\bm N}_1) ({\bm a}_{\lambda_2}\cdot{\bm N}_2)
({\bm a}_{\lambda_3}\cdot{\bm V})-4\nu_2({\bm a}_{\lambda_1}\cdot{\bm N}_1)
({\bm a}_{\lambda_3}\cdot{\bm N}_3)({\bm a}_{\lambda_2}\cdot{\bm V})
\nonumber\\
&&+({\bm a}_{\lambda_1}\cdot{\bm N}_1)\bigg\{
\frac{[\bm \rho\cdot\bm \rho]_{\lambda_2\lambda_3}}{2\nu_2\nu_3}+[{\bm \rho}\cdot
(\bm d_1-\bm d_2-\bm d_3 )]_{\lambda_2\lambda_3}-2\nu_2\nu_3[\bm d_1\cdot(\bm d_2+\bm d_3)]_{\lambda_2\lambda_3}\bigg\}
\nonumber\\
&&-({\bm a}_{\lambda_2}\cdot{\bm N}_2)\bigg\{
\frac{[\bm \rho\cdot\bm \rho]_{\lambda_1\lambda_3}}{2\nu_3}+[{\bm \rho}\cdot
(\bm d_1+\bm d_2-\bm d_3 )]_{\lambda_1\lambda_3}-2\nu_3[\bm d_3\cdot(\bm d_1+\bm d_2)]_{\lambda_1\lambda_3}\bigg\} \nonumber\\
&&-({\bm a}_{\lambda_3}\cdot{\bm N}_3)\bigg\{
\frac{[\bm \rho\cdot\bm \rho]_{\lambda_1\lambda_2}}{2\nu_2}+[{\bm \rho}\cdot
(\bm d_1-\bm d_2+\bm d_3) ]_{\lambda_1\lambda_2}-2\nu_2[\bm d_2\cdot(\bm d_1+\bm d_3)]_{\lambda_1\lambda_2}\bigg\}
\nonumber\\
&&+({\bm a}_{\lambda_1}\cdot{\bm \rho})[{\bm \rho}
\cdot\bm d_1]_{\lambda_2\lambda_3}-\frac{1}{\nu_2}({\bm a}_{\lambda_2}
\cdot{\bm \rho})[{\bm \rho}\cdot\bm d_3]_{\lambda_1\lambda_3}-
\frac{1}{\nu_3}({\bm a}_{\lambda_3}
\cdot{\bm \rho})[{\bm \rho}\cdot\bm d_2]_{\lambda_1\lambda_2}\nonumber\\
&&-\frac{\rho^2}{2}\bigg\{[{\bm a}_{\lambda_1}\cdot\bm d_1]_{\lambda_2\lambda_3}
-\frac{1}{\nu_2}[{\bm a}_{\lambda_2}\cdot\bm d_3]_{\lambda_1\lambda_3}
-\frac{1}{\nu_3}[{\bm a}_{\lambda_3}\cdot\bm d_2]_{\lambda_1\lambda_2}
\bigg\}\nonumber\\
&&+\frac{1}{\nu_2\nu_3}(\bm a_{\lambda_1}\cdot {\bm \rho})
(\bm a_{\lambda_2}\cdot {\bm \rho})(\bm a_{\lambda_3}\cdot {\bm \rho})
-\frac{\rho^2}{4\nu_2\nu_3}[\delta_{(1)}(\bm a_{\lambda_1}\cdot{\bm \rho})+\delta_{(2)}(\bm a_{\lambda_2}\cdot{\bm \rho})
+\delta_{(3)}(\bm a_{\lambda_3}\cdot{\bm \rho})]\, \nonumber\\
&&+\frac{2i}{ S}\left[\delta_{(1)}{\bm a}_{\lambda_1}\cdot({\bm N}_1-\bm V)+
\delta_{(2)}{\bm a}_{\lambda_2}\cdot({\bm N}_2+\nu_2\bm V)+
\delta_{(3)}{\bm a}_{\lambda_3}\cdot({\bm N}_3+\nu_3\bm V)\right]\, \nonumber\\
&&+U_1
-\nu_2\,U_2-\nu_3\,U_3+2\nu_2\nu_3\,U_4\Bigg>\,\, + \quad (2\leftrightarrow 3\, ,\,\, \bm \rho\rightarrow -\bm \rho )\,.
\end{eqnarray}
Here, we introduced the operator 
\begin{eqnarray}
[\bm X\cdot\bm Y]_{\lambda_i\lambda_j}\equiv (\bm X\cdot\bm a_{\lambda_i})
(\bm Y\cdot\bm a_{\lambda_j})+(\bm X\cdot\bm a_{\lambda_j})
(\bm Y\cdot\bm a_{\lambda_i})-(\bm X\cdot\bm Y)
(\bm a_{\lambda_i}\cdot\bm a_{\lambda_j})
\end{eqnarray}
for any two vectors $\bm X$ and $\bm Y$ and we used the abbreviations
\begin{eqnarray}\label{U}
&&U_1=-(\bm a_{\lambda_1}\cdot {\bm \rho})[\bm d_2\cdot \bm d_3]_{\lambda_2\lambda_3}+
(\bm a_{\lambda_2}\cdot {\bm \rho})[\bm d_2\cdot \bm d_3]_{\lambda_1\lambda_3}+
(\bm a_{\lambda_3}\cdot {\bm \rho})[\bm d_2\cdot \bm d_3]_{\lambda_1\lambda_2}\nonumber\\
&&+2(\bm a_{\lambda_1}\cdot{\bm \rho})(\bm a_{\lambda_2}\cdot \bm d_2)(\bm a_{\lambda_3}\cdot\bm d_3)
-(\bm d_2\cdot{\bm \rho})[\bm d_3\cdot\bm a_{\lambda_2}]_{\lambda_1\lambda_3}
-(\bm d_3\cdot{\bm \rho})[\bm d_2\cdot\bm a_{\lambda_3}]_{\lambda_1\lambda_2}\,,
\nonumber\\
&&\nonumber\\
&& U_2=(\bm a_{\lambda_1}\cdot {\bm \rho})[\bm d_1\cdot \bm d_2]_{\lambda_2\lambda_3}-
(\bm a_{\lambda_2}\cdot {\bm \rho})[\bm d_1\cdot \bm d_2]_{\lambda_1\lambda_3}+
(\bm a_{\lambda_3}\cdot {\bm \rho})[\bm d_1\cdot \bm d_2]_{\lambda_1\lambda_2}\nonumber\\
&&+2(\bm a_{\lambda_2}\cdot{\bm \rho})(\bm a_{\lambda_1}\cdot \bm d_2)(\bm a_{\lambda_3}\cdot\bm d_1)
-(\bm d_1\cdot{\bm \rho})[\bm d_2\cdot\bm a_{\lambda_3}]_{\lambda_1\lambda_2}
-(\bm d_2\cdot{\bm \rho})[\bm d_1\cdot\bm a_{\lambda_1}]_{\lambda_2\lambda_3}\,,\nonumber\\
&&\nonumber\\
&&
U_3=(\bm a_{\lambda_1}\cdot {\bm \rho})[\bm d_1\cdot \bm d_3]_{\lambda_2\lambda_3}+
(\bm a_{\lambda_2}\cdot {\bm \rho})[\bm d_1\cdot \bm d_3]_{\lambda_1\lambda_3}-
(\bm a_{\lambda_3}\cdot {\bm \rho})[\bm d_1\cdot \bm d_3]_{\lambda_1\lambda_2}\nonumber\\
&&+2(\bm a_{\lambda_3}\cdot{\bm \rho})(\bm a_{\lambda_1}\cdot \bm d_3)(\bm a_{\lambda_2}\cdot\bm d_1)
-(\bm d_1\cdot{\bm \rho})[\bm d_3\cdot\bm a_{\lambda_2}]_{\lambda_1\lambda_3}
-(\bm d_3\cdot{\bm \rho})[\bm d_1\cdot\bm a_{\lambda_1}]_{\lambda_2\lambda_3}\,,\nonumber\\
&&\nonumber\\
&& U_4=(\bm a_{\lambda_1}\cdot \bm d_2)[\bm d_1\cdot \bm d_3]_{\lambda_2\lambda_3}+
(\bm a_{\lambda_1}\cdot \bm d_3)[\bm d_1\cdot \bm d_2]_{\lambda_2\lambda_3}-
(\bm a_{\lambda_1}\cdot \bm d_1)[\bm d_2\cdot \bm d_3]_{\lambda_2\lambda_3}\nonumber\\
&&+2(\bm a_{\lambda_1}\cdot \bm d_1)(\bm a_{\lambda_2}\cdot \bm d_2)(\bm a_{\lambda_3}\cdot \bm d_3)
+(\bm d_1\cdot \bm d_2)[\bm d_3\cdot( \delta_{(2)}\bm a_{\lambda_2}-
\delta_{(3)}\bm a_{\lambda_3})] \nonumber\\
&&-(\bm d_1\cdot \bm d_3)[\bm d_2\cdot( \delta_{(2)}\bm a_{\lambda_2}-
\delta_{(3)}\bm a_{\lambda_3})]-(\bm d_2\cdot \bm d_3)[\bm d_1\cdot( \delta_{(2)}\bm a_{\lambda_2}+
\delta_{(3)}\bm a_{\lambda_3})]\, .
\end{eqnarray}
Equations (\ref{S2final}) and (\ref{S3final}) are valid for an arbitrary vector potential $\bm A(\phi)$ and they are our starting point for further consideration. The simplest case $\bm A(\phi)=-\bm E_0\phi$, where $\bm E_0$ is a constant vector perpendicular to $\bm\varkappa$, corresponds to the so-called crossed field with the electric field $\bm E=\bm E_0$ and the magnetic field $\bm H=\bm\varkappa\times\bm E_0$ , $|\bm H|=|\bm E|$. In this case there is no energy and momentum transfer from the field so that $\Omega=0$ and $\bm q= \bm 0$ (the momenta $\bm k_2$ and $\bm k_3$ of the final photons are parallel to the momentum $\bm k_1$ of the initial photon). Note that in this case the expansion of the amplitudes over $E_0/E_c$ starts with the terms of order of $(E_0/E_c)^3$  while the linear terms in $E_0/E_c$ vanish. We have checked that for a crossed field our amplitudes agree with those found in \cite{Papanyan74}. 

\section{Photon splitting in a monochromatic plane wave}

Below, we consider the external field to be a monochromatic plane wave with the frequency $\omega_0$. In this case, the helicity amplitude $M_{\sigma_1\sigma_2\sigma_3}$ has the form
\begin{equation}
M_{\sigma_1\sigma_2\sigma_3}=\sideset{}{'}\sum_{n=1}^\infty 2\pi\delta(\Omega-n\omega_0)M_{n,\,\sigma_1\sigma_2\sigma_3}
\end{equation}
after the integration over the variable $\phi$. In the above equation the prime means the summation over odd numbers $n$ and where $M_{n,\,\sigma_1\sigma_2\sigma_3}$ denotes the invariant amplitude of the photon splitting process with absorption of $n$ laser photons. Using the usual Fermi golden rule, we obtain for the photon splitting rate $d\dot W_n$ 
\begin{eqnarray}
\label{W}
d\dot W_{n,\,\sigma_1\sigma_2\sigma_3}= \frac{\pi}{\omega_1}|M_{n,\,\sigma_1\sigma_2\sigma_3}|^2
\frac{d\bm q}{\omega_1\Omega_n \cos\theta_0}=\frac{2\pi}{\omega_1}|M_{n,\,\sigma_1\sigma_2\sigma_3}|^2\,
d\nu_2\,d\phi_q\, ,
\end{eqnarray}
where $\Omega_n=n\omega_0$, $\cos\theta_0$ is defined in Eq. (\ref{nu}), and $\phi_q$ is the azimuth angle of the vector $\bm q$ in the plane perpendicular to $\bm \varkappa$. We remind the reader of the fact that $\bm q$ is a two-dimensional vector perpendicular to  $\bm \varkappa$ and $\bm k_1$.  Besides, ${d\bm q}/({\omega_1\Omega_n \cos\theta_0})$ coincides with the differential of the solid angle of the vector $\bm k_2$ in the frame where $\omega_1=\Omega_n$. It is also useful to express the rate $d\dot W_{n,\,\sigma_1\sigma_2\sigma_3}$  in terms of the polar angle $\theta_2=\angle (\bm k_2, \bm \varkappa) $ (or of the polar angle $\theta_3=\angle (\bm k_3, \bm \varkappa) $). Using the expressions (\ref{theta0}) valid for any $\Omega$ we obtain the relations
\begin{equation}\label{theta23}
\nu_{2,3}=\frac{1}{1+\dfrac{\omega_1}{\Omega}\cot^2(\theta_{2,3}/2)}\, ,  \quad \cot(\theta_2/2) \cot(\theta_3/2)=\frac{\Omega}{\omega_1}\,.
\end{equation}
As a result, we find from Eq. (\ref{W}) that
\begin{eqnarray}
\label{W1}
d\dot W_{n,\,\sigma_1\sigma_2\sigma_3}=\frac{\pi}{\Omega_n}\,|M_{n,\,\sigma_1\sigma_2\sigma_3}|^2\,
\frac{\sin\theta_2\,d\theta_2\,d\phi_q}{\left[\sin^2(\theta_2/2)+\dfrac{\omega_1}{\Omega_n}\cos^2(\theta_2/2)\right]^2}\, .
\end{eqnarray}

In the general case of an elliptically polarized and monochromatic laser wave, the vector potential reads
\begin{equation}
\bm A(\phi)=\bm A_1\cos(\omega_0\phi)+\bm A_2\sin(\omega_0\phi)\, ,
\end{equation}
where $\bm A_1\cdot\bm A_2=\bm A_1\cdot\bm\varkappa=\bm A_2\cdot\bm\varkappa=0$. Generally speaking, the vectors $\bm A_1$
and $\bm A_2$ are not parallel to the unit vectors $\bm a_1$ and $\bm a_2$ introduced in Eq. (\ref{varepsilon}), respectively, as there is an azimuth asymmetry of the photon splitting amplitude for an elliptically polarized laser wave (we point out that $\bm a_1$ has been chosen to be parallel to $\bm q$). However, in a circularly polarized laser field it is $|\bm A_1|=|\bm A_2|$, then the photon splitting rate is symmetric with respect to the azimuth angle $\phi_q$ of the vector $\bm q$, and the vector $\bm A_1$ can be directed along $\bm a_1$. 

The photon splitting rate can be converted into a cross section according to the relation
\begin{eqnarray}
\label{flux}
d\sigma = \frac{d\dot W}{2\Phi}\,,\quad \Phi = \frac{1}{8\pi}\omega_0(A_1^2+A_2^2)=\frac{\omega_0m^2}{8\pi\alpha}(\xi_1^2+\xi_2^2)\, ,
\end{eqnarray}
with $\Phi$ being the laser photon flux and $\xi_{1,2}=eA_{1,2}/m$. Note that this cross section has a restricted meaning because in the presence of a strong field it still depends on the laser flux through the parameters $\xi_{1,2}$. In terms of the parameters $\xi_{1,2}$ and $\eta$, the parameter $\chi$ introduced in Eq. (\ref{chieta}) has the form $\chi=\eta\sqrt{(\xi_1^2+\xi_2^2)/2}$. 

The integral over $\phi$ in Eqs. (\ref{S2final}) and (\ref{S3final}) can be taken analytically for an elliptically polarized laser wave. However,
the result is very cumbersome even in the case of linear laser polarization. This is why we limit ourselves to the case of a circularly polarized laser field. In this case the answer is essentially simpler (but not simple). Thus we set $|\bm A_1|=|\bm A_2|=A$, $\xi_{1}=\xi_2=\xi=eA/m$ and direct $\bm A_1$ along $\bm a_1$ and $\bm A_2$ along $\bm a_2$ (for positive helicity (p.\,h.)) or $\bm A_2$ along $-\bm a_2$ (for negative helicity (n.\,h.)).  Only the case of positive helicity of the laser field will be considered below and the amplitudes for negative helicity of the laser field polarization can be obtained by means of the relation $M_{\sigma_1\sigma_2 \sigma_3}(\mbox{n.\,h.})=-M_{\bar\sigma_1\bar\sigma_2\bar\sigma_3}(\mbox{p.\,h.})$, where $\bar\sigma_i$ denotes the helicity opposite to $\sigma_i$. 

Now, by using the substitution rule of Eqs. (\ref{subs}) and by carrying out the change of variables $s_1=us$ and $s_2=(1-u)s$ in Eq. (\ref{S2final}), we obtain for the nonzero contributions of $M^{(2)}_{n,\sigma_1\sigma_2 \sigma_3}$ the following result
\begin{eqnarray}\label{M2circular}
&&M^{(2)}_{n,++-}=G(1)+G(-\nu_3)\, ,\,\, M^{(2)}_{n,+-+}=G(1)+G(-\nu_2)\,,\,\,  \quad M^{(2)}_{n,---}=G(-\nu_2)+G(-\nu_3)\,,\nonumber\\
&&G(\nu)=\delta_{n1}\,\frac{m\xi e^3}{4\sqrt{2}\pi^2}\int_0^\infty\frac{ ds}{s}\int_0^1du
\exp\left\{-is\left[1+\xi^2\left(1-\frac{\sin^2\vartheta}{\vartheta^2}\right)\right]\right\}\nonumber\\
&&\times\mbox{e}^{i\vartheta}\left[2i\left(\mbox{e}^{i\vartheta} -   \frac{\sin\vartheta}{\vartheta}\right)+
\,\frac{\sin\vartheta}{u(1-u)}\right]\,;\quad \vartheta=2u(1-u)s\eta\nu \, .
\end{eqnarray}
 
In order to write $M^{(3)}_{n,\sigma_1\sigma_2 \sigma_3}$ in a compact form, we introduce the following definitions
\begin{eqnarray}\label{not}
&&f_j=\sin(2\eta s_j\tau_j)\mbox{e}^{i2\eta s_j\tau_j}\,,\quad 
\zeta=\frac{i}{\tau_1\eta}\left(\frac{f_2}{s_2\tau_2}-\frac{f_3}{s_3\tau_3}\right)\,,\quad \phi_0=\arg{\zeta}\,,\nonumber\\
&&F_0=\frac{\xi^2}{4\eta^2 S\tau_1\tau_2\tau_3}\left(\frac{\nu_2\nu_3}{\tau_1}|f_1|^2+\frac{\nu_2}{\tau_2}|f_2|^2
-\frac{\nu_3}{\tau_3}|f_3|^2\right)-\xi^2 S\, ,\nonumber\\
&& Z=\xi|\zeta|\frac{\rho_n s_2s_3}{S}\,,\quad \rho_n=2\sqrt{n\eta\nu_2\nu_3}\,,\quad  S=s_1+s_2+s_3\, .
\end{eqnarray}
In terms of these functions, the contribution $M^{(3)}_{n,\sigma_1\sigma_2 \sigma_3}$ has the form
\begin{eqnarray}\label{M3circular}
M^{(3)}_{n,\sigma_1\sigma_2 \sigma_3}&=&\frac{me^3}{4\sqrt{2}\pi^2}\int\negthickspace\negthickspace\int\limits_0^{\infty}
\negthickspace\negthickspace\int
\frac{ds_1\,ds_2\,ds_3}{ S^2}\exp\left[i\left(4n\eta\frac{s_2s_3}{ S}+F_0- S\right)\right]\nonumber\\
&&\times \sum_{j=-3}^{3}\, \mbox{e}^{-i(n+j)\phi_0}\,J_{n+j}(Z)B_{j,\sigma_1\sigma_2 \sigma_3}\, ,
\end{eqnarray}
where $J_l(Z)$ are ordinary Bessel functions. The coefficients $B_{j,\sigma_1\sigma_2 \sigma_3}$ are expressed via the functions $f_j$ introduced in Eq. (\ref{not})
and they are presented in the Appendix. The expressions (\ref{M2circular}) and (\ref{M3circular}) are valid for any value of the parameters $\eta$ and $\xi=\chi/\eta$. As it is clear from Eq. (\ref{M3circular}), the photon splitting amplitudes are not equal to zero for any number of absorbed laser photons. This is because the angle between the two final photons is not zero ($\rho_n\neq 0$) and the conservation of the projection $J_z$ of the total angular momentum does not imply any selection rule. Note that the expansion of the amplitude $M_{n,\sigma_1\sigma_2 \sigma_3}=M^{(2)}_{n,\sigma_1\sigma_2 \sigma_3}+M^{(3)}_{n,\sigma_1\sigma_2 \sigma_3}$ for a fixed $n$ (see Eqs. (\ref{M2circular}) and (\ref{M3circular})), contains in general all odd powers $k$ of the parameter $\chi$ (or, equivalently, $\xi$) starting from $k=n$. The terms with $k>n$ in $M_{n,\sigma_1\sigma_2 \sigma_3}$ correspond to rescattering processes of absorption and emission of laser photons with a net absorption of $n$ laser photons.

In the next Section we consider the asymptotics of the amplitudes $M^{(2)}_{n,\sigma_1\sigma_2 \sigma_3}$ and  $M^{(3)}_{n,\sigma_1\sigma_2 \sigma_3}$ in some limiting cases.

\section{Asymptotics of the amplitudes}
Various limiting cases of the photon splitting amplitudes for a monochromatic plane wave are studied in this Section.

\subsection{Amplitudes for small $\eta$ and small $\chi$}
We first consider the case of both small $\eta$ and $\chi$, $\eta\ll 1$ and $\chi\ll 1$. Then, the leading contribution to the amplitudes 
is determined by the ratio $\eta/\chi^2$ \cite{Aff87}. For $\eta/\chi^2\gg 1$, the nonzero amplitudes in the leading approximation are given by 
the linear terms in $\chi$ in Eqs.(\ref{M2circular}) and (\ref{M3circular}) and correspond to $n=1$. We obtain
\begin{eqnarray}\label{smallchi}
&&M_{1,+++}=11{\cal N}\,,\quad M_{1,+--}=-2(1-\nu_2\nu_3){\cal N}\,,\quad M_{1,--+}=11\nu_2^2{\cal N}\,,\nonumber\\
&& M_{1,-+-}= 11\nu_3^2{\cal N} \,,  \quad  {\cal N}=-\frac{ime^3\eta\chi}{45\sqrt{2}\pi^2}\, .
\end{eqnarray}
These amplitudes are nothing but the usual photon-photon scattering amplitudes for small Mandelstam variables 
$s=4\eta m^2$, $t=-4\eta\nu_2 m^2$ and $u=-4\eta\nu_3 m^2$ (see, e.g., \cite{LL}). For $\eta/\chi^2\ll 1$, the leading contributions are proportional to $\chi^3$ and correspond to both $n=1$ and $n=3$. The nonzero amplitudes in the leading approximation are given by
\begin{eqnarray}\label{n1}
&&M_{1,+-+}=M_{1,++-}=M_{1,---}=-i\frac{37}{15}{\widetilde{\cal N}}\, ,\nonumber\\
&& M_{3,+--}=i{\widetilde{\cal N}}\, ,\quad {\widetilde{\cal N}}=\frac{2\sqrt{2}me^3\chi^3\nu_2\nu_3}{21\pi^2}\, . 
\end{eqnarray}
These results are in agreement with those obtained in \cite{Aff87}.

\subsection{Amplitudes for small $\eta$ and fixed $\chi$}

\begin{figure}
\begin{center}
\includegraphics[width=10cm]{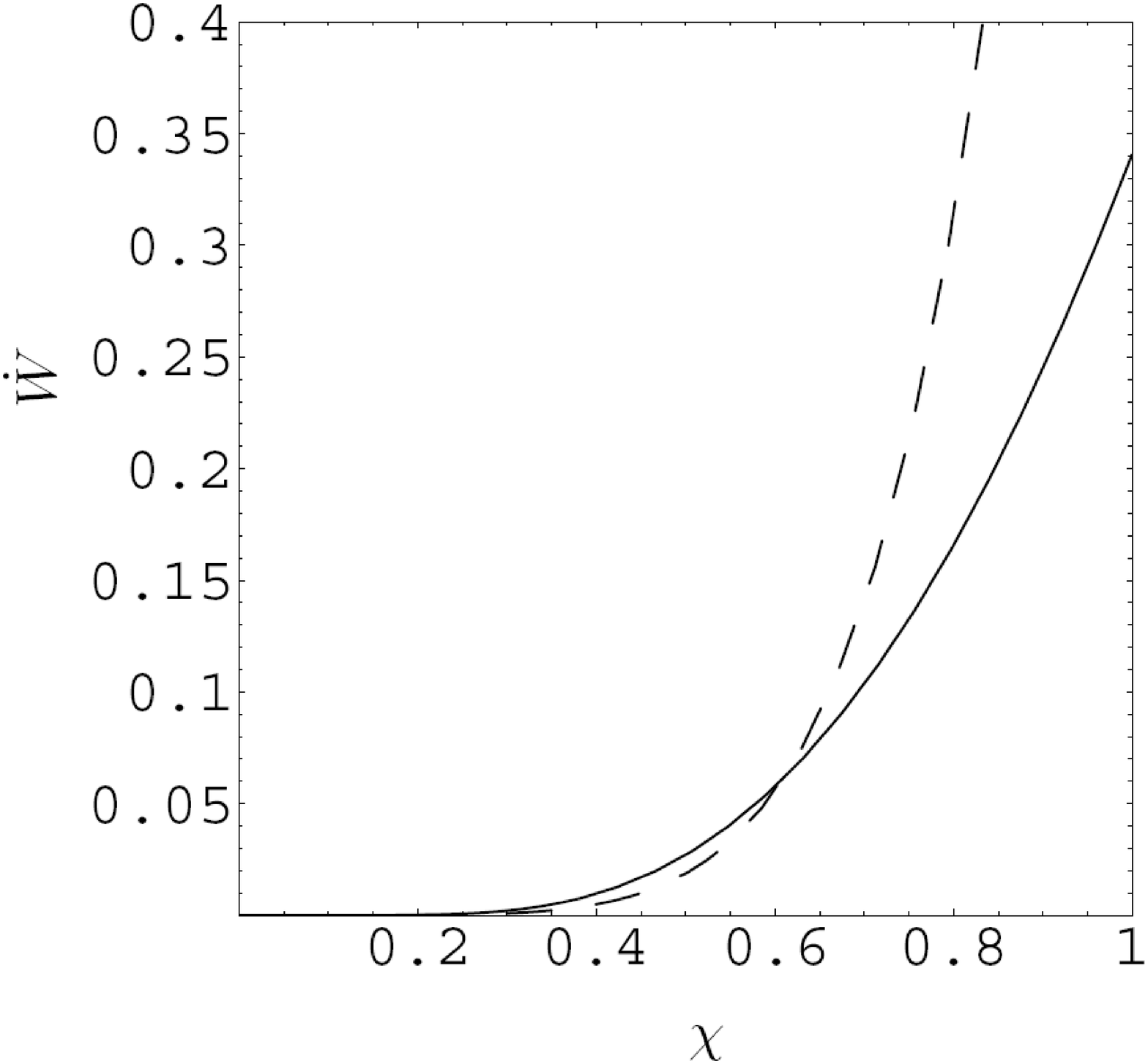}
\end{center}
\caption{The photon splitting rate $\dot W$ in a circularly polarized laser field in  units $(\pi^2m^2\alpha^3/\omega_1)\times 10^{-4}$ as a function of $\chi$ at $\eta\ll 1$. The solid line is obtained by using the amplitudes (\ref{M3as}), the dashed line is based on the small-$\chi$ asymptotics (\ref{n1}). The rate is averaged over the polarization of the initial photon, summed up over the polarizations of the final ones and integrated over $\nu_2$ and $\phi_q$.}\label{fig:chi}
\end{figure}

We proceed now to the region of parameters $\eta\ll 1$ and $\chi$ fixed. In this case the leading terms of the amplitudes in a circularly polarized laser field contain all powers of $\chi$ but only for $n=1$ and $n=3$ they still are not equal to zero. The reason is the following: at $\eta\ll 1$, in a frame where $\omega_1\gg\omega_0$, the vectors $\bm k_2$ and  $\bm k_3$ are almost parallel to $\bm k_1$ and anti-parallel to $\bm\varkappa$. Since the clockwise polarized laser field contains photons with projection of the total angular momentum $J_z=+1$ then, due to the conservation of $J_z$, only the values $n=1$ and $n=3$ are allowed and only the amplitudes $M_{1,++-}$, $M_{1,+-+}$, $M_{1,---}$ and $M_{3,+--}$ are different zero. For these amplitudes we found in the leading approximation with respect to $\eta$: 
\begin{eqnarray}\label{M3as}
&&M_{1,++-}=\frac{ime^3\chi^3}{\sqrt{2}\pi^2}\int\negthickspace\negthickspace\int\limits_0^{\infty}
\negthickspace\negthickspace\int
{ds_1\,ds_2\,ds_3}\,\mbox{e}^{-i(S+\psi_0)}\Big\{-\frac{i(z_1+z_3)}{S^2\chi^2}\nonumber\\
&&+S\left[z_1z_2z_3-\nu_2\nu_3z_1+\nu_2z_3+\nu_3z_2t_3^2-\nu_2\nu_3(1+t_3^2)\right]\Big\}
+{\cal G}(1)+{\cal G}(-\nu_3)\,; \nonumber\\
&&{}\nonumber\\
&&M_{1,---}=\frac{ime^3\chi^3}{\sqrt{2}\pi^2}\int\negthickspace\negthickspace\int\limits_0^{\infty}
\negthickspace\negthickspace\int
{ds_1\,ds_2\,ds_3}\,\mbox{e}^{-i(S+\psi_0)}\Big\{-\frac{i(z_2+z_3)}{S^2\chi^2}\nonumber\\
&&+ S\left[z_1z_2z_3-\nu_2\nu_3z_1t_1^2+\nu_2z_3+\nu_3z_2-\nu_2\nu_3(1+t_1^2)\right]\Big\}
+{\cal G}(-\nu_2)+{\cal G}(-\nu_3)\, ;\nonumber\\
&&{}\nonumber\\
&&M_{3,+--}=\frac{ime^3\chi^3}{\sqrt{2}\pi^2}\int\negthickspace\negthickspace\int\limits_0^{\infty}
\negthickspace\negthickspace\int
{ds_1\,ds_2\,ds_3}\,\mbox{e}^{-i(S+\psi_0)}\,\nonumber\\
&&\times S\left[z_1z_2z_3-\nu_2\nu_3z_1t_1^2+\nu_2z_3t_2^2+\nu_3z_2t_3^2+2\nu_2\nu_3t_1t_2t_3\right]\,.
\end{eqnarray}
Here we introduced the following quantities:
\begin{eqnarray}\label{psi}
&&{\cal G}(\nu)=\nu\frac{me^3\chi}{2\sqrt{2}\pi^2}\int_0^\infty ds\int_0^1 du\,[1-2u(1-u)]\mbox{e}^{-i(s+\psi_1)}\, ;\nonumber\\
&&z_1=2x_1(x_2\nu_2+x_3\nu_3)+2x_2x_3-1\, ,\quad z_2=2x_1(-x_2\nu_2+x_3\nu_3)-2x_2x_3+\nu_2\, ,\nonumber\\
&&z_3=2x_1(x_2\nu_2-x_3\nu_3)-2x_2x_3+\nu_3\,,\quad   x_j=\frac{s_j}{ S}\,,\quad t_j=1-2x_j\,,\nonumber\\
&&\psi_0=\frac{4}{3}\chi^2S^3[x_2^2x_3^2+2x_1x_2x_3(\nu_2x_2+\nu_3x_3)+x_1^2(\nu_2x_2-\nu_3x_3)^2]\,,\nonumber\\
&&\psi_1=\frac{4}{3}\nu^2\chi^2u^2(1-u)^2s^3\,.
\end{eqnarray}
The amplitude $M_{1,+-+}$ is obtained from  $M_{1,++-}$ by the replacement $\nu_2\leftrightarrow\nu_3$. Equations (\ref{M3as}) are independent of the parameter $\eta$ and are valid for arbitrary values of $\chi$. The linear terms of the expansion of these amplitudes with respect to $\chi$ vanish. Then, the expansion starts with the terms proportional to $\chi^3$ that coincide with Eqs. (\ref{n1}). As it is known, both the electromagnetic invariants $F_{\mu\nu}F^{\mu\nu}$ and $\epsilon_{\mu\nu\rho\sigma}F^{\mu\nu}F^{\rho\sigma}$ are equal to zero for a plane wave, $F^{\mu\nu}$ being the tensor of the electromagnetic field. The Euler-Heisenberg Lagrangian depends on these two invariants but not on $\chi$ which in our case has the form $\chi=\sqrt{-(eF_{\mu\nu}k_1^{\nu})^2}/(2m^3)$. Therefore only the terms (\ref{n1}) can be obtained by means of this effective Lagrangian
but not the total answer (\ref{M3as}). Thus, the applicability of the Euler-Heisenberg Lagrangian to the calculation of the
photon splitting amplitudes is restricted by the conditions $\eta\ll 1$ and $\chi\ll 1$. It is worth noting that the region of applicability of the small-$\chi$ asymptotics is very narrow. For instance, at $\chi=0.1$ the difference between the leading terms (\ref{n1}) and the exact result (\ref{M3as}) is already around $20 \%$. We illustrate the importance of high-order terms in $\chi$ in Fig. \ref{fig:chi} where we show the total photon splitting rate $\dot W$ at intermediate values of the parameter $\chi$. The solid line is obtained by using the amplitudes (\ref{M3as}), the dashed line by employing Eq. (\ref{n1}).

\begin{figure}
\begin{center}
\includegraphics[width=10cm]{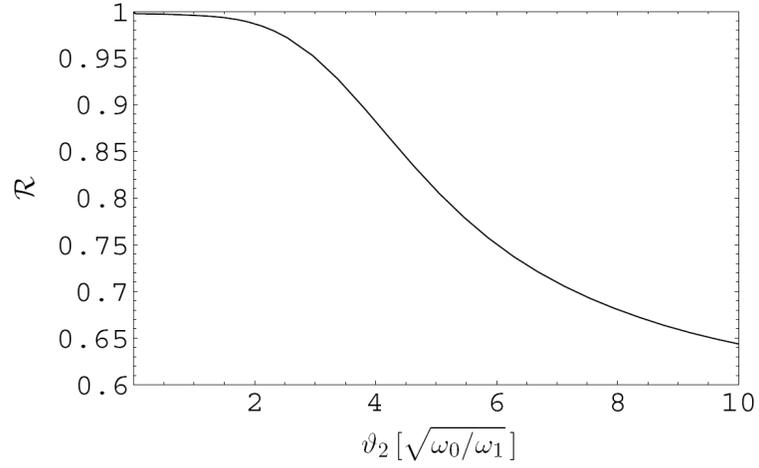}
\end{center}
\caption{The ratio $\cal R$ defined in Eq. (\ref{R}) calculated at $\chi=1$ with the help of Eq. (\ref{M3as}) as a function of the angle $\vartheta_2$, measured in units of $\sqrt{\omega_0/\omega_1}$.}\label{angle}
\end{figure}

At $\omega_1\gg m$ it is very difficult to measure the energy of the final photons with the accuracy necessary to distinguish between the process of photon splitting with one-photon absorption and three-photon absorption. However, there is another possibility to distinguish these two processes which is based on the use of the angular distribution of the final photons. At $\omega_1\gg\omega_0$, the final photons are almost parallel to the initial one so that $\vartheta_{2,3}\ll 1$ where  $\vartheta_{2}=\angle (\bm k_2, \bm k_1)$ and $\vartheta_{3}=\angle (\bm k_3, \bm k_1) $. It follows from  Eq. (\ref{W1}) that
\begin{eqnarray}
\label{Wvaromega}
d\dot W_{n,\,\sigma_1\sigma_2\sigma_3}  = \frac{\pi}{n\omega_0}\,|M_{n,\,\sigma_1\sigma_2\sigma_3}|^2\,
\frac{\vartheta_2\,d\vartheta_2\,d\phi_q}{\left(1+\dfrac{\omega_1}{4n\omega_0}\vartheta_2^2\right)^2}\, .
\end{eqnarray}
Using the expressions (\ref{M3as}), we have tabulated the ratio 
\begin{equation}\label{R}
{\cal R}=\frac{\sum_{\sigma_j}d\dot W_{1,\,\sigma_1\sigma_2\sigma_3}}
{\sum_{\sigma_j}(d\dot W_{1,\,\sigma_1\sigma_2\sigma_3}+ d\dot W_{3,\,\sigma_1\sigma_2\sigma_3})}\, .
\end{equation}
The result of this tabulation at $\chi=1$ is shown in Fig. \ref{angle}. It can be seen that $\cal R$ noticeably differs from unity at angles $\vartheta_2$ such that $\vartheta_2\gg\sqrt{\omega_0/\omega_1}$. 

It is interesting to compare the amplitudes of photon splitting in a circularly polarized laser field with those in a linearly polarized field.
In the latter case for  $\eta\ll 1$ and fixed $\chi$, the leading terms of the amplitudes are not zero for all odd $n$. This is due to the fact that a linearly polarized laser beam contains photons with both $J_z=+1$ and $J_z=-1$. In this sense the situation reminds us of the properties of the polarization operator in the laser field \cite{BMS75}. In the case of a linearly polarized laser field it is convenient to perform the calculations of the amplitudes in terms of the linear polarization of photons. Since the vectors $\bm k_2$ and  $\bm k_3$ are almost parallel to $\bm k_1$, the azimuth asymmetry disappears. Therefore, we can choose $\bm A(\phi)=\bm a_1 A_1\cos(\omega_0\phi)$ and we have $\chi=\eta\xi_1/\sqrt{2}$ with $\xi_1=eA_1/m$. The polarization of each photon is parallel either to $\bm a_1$ or to $\bm a_2$ so that our amplitudes coincide with the coefficients $R_{\lambda_1\lambda_2\lambda_3}$ in Eq. (\ref{RVM}). Note that $\bm a_2=\bm\varkappa\times\bm a_1$ and thus $\bm a_2$ is a pseudo-vector. Therefore, the nonzero amplitudes in the collinear approximation are those which either do not contain or contain twice the indices $\lambda=2$, namely, $M_{n,111}$, $M_{n,122}$,  $M_{n,212}$, and  $M_{n,221}$. Putting $\rho_n\propto\sqrt{\eta}$ equal to zero in Eqs. (\ref{S2final}) and (\ref{S3final}) and performing the integral over $\phi$, we obtain the leading contributions of the amplitudes at $\eta\ll 1$ for a linearly polarized laser field
\begin{eqnarray}\label{ML}
&&M_{n,111}=\frac{i^{\frac{n+3}{2}}\sqrt{2}me^3\chi^3}{\pi^2}\int\negthickspace\negthickspace\int\limits_0^{\infty}
\negthickspace\negthickspace\int
{ds_1\,ds_2\,ds_3}\,\mbox{e}^{-i(S+\psi_0)}\bigg\{-\frac{i(z_1+z_2+z_3){\cal D}_n^{(1)}(\psi_0)}{2S^2\chi^2}\nonumber\\
&&+S{\cal D}_n^{(2)}(\psi_0)\,\big[2z_1z_2z_3-\nu_2\nu_3(1+t_1^2)z_1+\nu_2(1+t_2^2)z_3+\nu_3(1+t_3^2)z_2\nonumber\\
&&-4\nu_2\nu_3+\nu_2\nu_3(1+t_1)(1+t_2)(1+t_3)\big]\bigg\}
+{\widetilde{\cal G}}_n(1)+{\widetilde{\cal G}}_n(-\nu_2)+{\widetilde{\cal G}}_n(-\nu_3)    \,; \nonumber\\
&&{}\nonumber\\
&&M_{n,122}=\frac{i^{\frac{n+3}{2}}\sqrt{2}me^3\chi^3}{\pi^2}\int\negthickspace\negthickspace\int\limits_0^{\infty}
\negthickspace\negthickspace\int
{ds_1\,ds_2\,ds_3}\,\mbox{e}^{-i(S+\psi_0)}\bigg\{-\frac{iz_1{\cal D}_n^{(1)}(\psi_0)}{2S^2\chi^2}\nonumber\\
&&-S{\cal D}_n^{(2)}(\psi_0)\nu_2\nu_3(1-t_1)\big[(1+t_1)z_1+1-t_2t_3\big]\bigg\}
+{\widetilde{\cal G}}_n(1)    \,; \nonumber\\
&&{}\nonumber\\
&&M_{n,212}=\frac{i^{\frac{n+3}{2}}\sqrt{2}me^3\chi^3}{\pi^2}\int\negthickspace\negthickspace\int\limits_0^{\infty}
\negthickspace\negthickspace\int
{ds_1\,ds_2\,ds_3}\,\mbox{e}^{-i(S+\psi_0)}\bigg\{-\frac{iz_2{\cal D}_n^{(1)}(\psi_0)}{2S^2\chi^2}\nonumber\\
&&+S{\cal D}_n^{(2)}(\psi_0)\nu_3(1-t_3)\big[(1+t_3)z_2-\nu_2(1-t_1t_2)\big]\bigg\}
+{\widetilde{\cal G}}_n(-\nu_2)    \,,
\end{eqnarray}
where 
\begin{eqnarray}\label{Gl}
&&{\cal D}_n^{(1)}(y)=J_{\frac{n-1}{2}}(y)-iJ_{\frac{n+1}{2}}(y)\,,\nonumber\\
&&{\cal D}_n^{(2)}(y)=\left(1+i\frac{n-1}{4y}\right)J_{\frac{n-1}{2}}(y)
-i\left(1-i\frac{n+1}{4y}\right)J_{\frac{n+1}{2}}(y)\,,\nonumber\\
&&{\widetilde{\cal G}}_n(\nu)=\nu\frac{i^{\frac{n+1}{2}}me^3\chi}{2\sqrt{2}\pi^2}\int_0^\infty ds\int_0^1 du\,[1-2u(1-u)]
\mbox{e}^{-i(s+\psi_1)}{\cal D}_n^{(1)}(\psi_1)\,, \nonumber\\
\end{eqnarray}
and the other notations are presented in Eq. (\ref{psi}). The amplitudes $M_{n,221}$ are obtained from $M_{n,212}$ by the replacement $\nu_2\leftrightarrow\nu_3$. In Fig. 4 we show the photon splitting rate $\dot{W}$ in a linearly polarized laser field as a function of the number $n$ of laser photons absorbed. The rate has been calculated from the above amplitudes (\ref{ML}) with $\chi=1$. The figure shows that, as the number $n$ increases, the photon splitting rate decreases quite rapidly and in a non-perturbative way.

The leading contributions to the amplitudes at $\chi\ll 1$ have the form [see also Eq. (\ref{n1})]
\begin{eqnarray}\label{n1l}
&&M_{1,111}=\frac{24}{5}{\widetilde{\cal N}}\,,\quad M_{3,111}=-\frac{8}{5} {\widetilde{\cal N}}\, , \nonumber\\
&&M_{1,122}=M_{1,212}=\frac{13}{5}{\widetilde{\cal N}}\,\quad M_{3,122}=M_{3,212}=-\frac{13}{15}{\widetilde{\cal N}}\,.
\end{eqnarray}

Note that at $\eta\ll 1$ there is a simple relation between the amplitudes of the photon splitting in a crossed field and those calculated in a linearly polarized laser field \cite{Papanyan74}. The amplitudes in a crossed field depend only on the parameter $\chi$ while those in a laser field depend only on $\chi$ in the leading approximation for $\eta\ll 1$. In this case, it can easily be shown that 
\begin{equation}\label{Mn}
M_{n,\lambda_1\lambda_2\lambda_3}^{(\text{lin})}(\chi)=\int_0^{2\pi}\frac{d\phi}{2\pi}\mbox{e}^{in\phi}
M_{\lambda_1\lambda_2\lambda_3}^{(\text{cross})}(\chi\sin\phi)\,.
\end{equation}
Therefore, due to Parseval's identity, we have
\begin{equation}
\sum_n|M_{n,\lambda_1\lambda_2\lambda_3}^{(\text{lin})}(\chi)|^2=\int_0^{2\pi}\frac{d\phi}{2\pi}
|M_{\lambda_1\lambda_2\lambda_3}^{(\text{cross})}(\chi\sin\phi)|^2\,.
\end{equation}
For the same reason, in the case of a circularly polarized laser field the following relation holds
\begin{equation}\label{circRitus}
\sum_{n,\,\sigma_i}|M_{n,\sigma_1\sigma_2\sigma_3}^{(\text{circ})}(\chi)|^2=\sum_{\sigma_i}
|M_{\sigma_1\sigma_2\sigma_3}^{(\text{cross})}(\chi)|^2\,,
\end{equation}
where the indices $\sigma_i$ denote the helicities of the external photons. We emphasize that Eq. (\ref{circRitus}) is not valid without the summation over $\sigma_i$. This is because the rotational symmetry around the vector $\bm \varkappa$ is absent in a crossed field (see discussion below Eq. (\ref{U})), as is the conservation of the projection of the total angular momentum $J_z$. Therefore, it is impossible to obtain the amplitudes $M_{n,\sigma_1\sigma_2\sigma_3}^{(\text{circ})}(\chi)$ in a circularly polarized field from those in a crossed field by using any Fourier transformation as it was done in Eq. (\ref{Mn}). 

Now, at $\chi\gg 1$ the asymptotics (\ref{M3as}) obtained for the circularly polarized field read

\begin{eqnarray}\label{Mclarge}
&&M_{1,++-}={\cal N}_c\Bigg\{ \int_0^1du\,u\int_0^1\frac{dw}{{\tilde \psi}^{4/3}}
\big[4(z_1+z_3){\tilde \psi}+z_1z_2z_3-\nu_2\nu_3z_1\nonumber\\
&&+\nu_2z_3+\nu_3z_2t_3^2-\nu_2\nu_3(1+t_3^2)\big]+\frac{8\Gamma^2(1/3)}{5\Gamma(2/3)}(1-\nu_3^{1/3})\Bigg\}\,; \nonumber\\
&&{}\nonumber\\
&&M_{1,---}={\cal N}_c\Bigg\{ \int_0^1du\,u\int_0^1\frac{dw}{{\tilde \psi}^{4/3}}\big[ 4(z_2+z_3){\tilde \psi}
+ z_1z_2z_3-\nu_2\nu_3z_1t_1^2        \nonumber\\
&&+\nu_2z_3+\nu_3z_2-\nu_2\nu_3(1+t_1^2)\big]
-\frac{8\Gamma^2(1/3)}{5\Gamma(2/3)}(\nu_2^{1/3}+\nu_3^{1/3})\Bigg\}\,;  \nonumber\\
&&{}\nonumber\\
&&M_{3,+--}={\cal N}_c\int_0^1du\,u\int_0^1\frac{dw}{{\tilde \psi}^{4/3}}
\big(z_1z_2z_3-\nu_2\nu_3z_1t_1^2+\nu_2z_3t_2^2+\nu_3z_2t_3^2+2\nu_2\nu_3t_1t_2t_3\big)\,,
\end{eqnarray}
where
\begin{eqnarray}\label{psi1}
&&{\cal N}_c=\frac{me^3(6\chi)^{1/3}\mbox{e}^{-i\pi/6}\Gamma(1/3)}{24\sqrt{2}\pi^2}\, ;\nonumber\\
&&{\tilde\psi}=x_2^2x_3^2+2x_1x_2x_3(\nu_2x_2+\nu_3x_3)+x_1^2(\nu_2x_2-\nu_3x_3)^2\,,\nonumber\\
&&x_1=1-u\, ,\quad x_2=uw\, ,\quad x_3=u(1-w)\, ,
\end{eqnarray}
and $z_j$ and $t_j$ are expressed via $x_j$ as in Eq. (\ref{psi}). Analogously, at $\chi\gg 1$ we find from Eq. (\ref{ML}) for a linearly polarized laser field
\begin{eqnarray}\label{MLlarge}
&&M_{n,111}={\cal N}_n\Bigg\langle \int_0^1du\,u\int_0^1\frac{dw}{{\tilde \psi}^{4/3}}
\big[4(z_1+z_2+z_3){\tilde \psi}
+2z_1z_2z_3-\nu_2\nu_3(1+t_1^2)z_1+\nu_2(1+t_2^2)z_3\nonumber\\
&&+\nu_3(1+t_3^2)z_2-4\nu_2\nu_3+\nu_2\nu_3(1+t_1)(1+t_2)(1+t_3)\big]
+\frac{8\Gamma^2(1/3)}{5\Gamma(2/3)}(1-\nu_2^{1/3}-\nu_3^{1/3})\Bigg\rangle  \,; \nonumber\\
&&{}\nonumber\\
&&M_{n,122}={\cal N}_n\Bigg\langle \int_0^1du\,u\int_0^1\frac{dw}{{\tilde \psi}^{4/3}}
\big\{4z_1{\tilde\psi}-\nu_2\nu_3(1-t_1)\big[(1+t_1)z_1+1-t_2t_3\big]\big\}
+\frac{8\Gamma^2(1/3)}{5\Gamma(2/3)}\Bigg\rangle   \,; \nonumber\\
&&{}\nonumber\\
&&M_{n,212}={\cal N}_n\Bigg\langle \int_0^1du\,u\int_0^1\frac{dw}{{\tilde \psi}^{4/3}}\big\{4z_2{\tilde\psi}
+\nu_3(1-t_3)\big[(1+t_3)z_2-\nu_2(1-t_1t_2)\big]\big\}
\nonumber\\
&&-\frac{8\Gamma^2(1/3)}{5\Gamma(2/3)}\nu_2^{1/3}\Bigg\rangle    \,,
\quad {\cal N}_n=\frac{me^3(6\chi)^{1/3}\mbox{e}^{i\pi/3}\Gamma(1/6)\Gamma(n/2-1/6)}{2^{5/6}72 \pi^{5/2}\Gamma(n/2+7/6)}\, .
\end{eqnarray}
The dependence on $n$ in these expressions is contained in the coefficients ${\cal N}_n$ which at large $n$ rapidly decrease as
$n^{-4/3}$. Instead, the dependence of the amplitudes on the parameter $\chi$, both for a linearly and a circularly polarized laser field, is weak and proportional to $\chi^{1/3}$.

\subsection{Amplitudes for large $\eta$ and fixed $\xi=\chi/\eta$}
The structure of the photon splitting amplitudes at large $\eta$ and fixed $\xi=\chi/\eta$ is rather interesting because the function $G(\nu)$ introduced in Eq. (\ref{M2circular}) and which enters the contributions $M^{(2)}_{n,\sigma_1\sigma_2\sigma_3}$ has the form
\begin{equation}
\begin{split}
G(\nu)&=\delta_{n1}\,\frac{im\xi e^3}{8\sqrt{2}\pi^2}\left\{\log^2\left[\frac{4\eta(-\nu-i0)}{1+\xi^2}\right]-4\right.\\
&\left.-4i\lambda\int_0^\infty d\vartheta\frac{\sin\vartheta\, \mbox{e}^{i\lambda\vartheta}}{\vartheta}\log\left[\frac{1+\xi^2}{1+\xi^2(1-\sin^2\vartheta/\vartheta^2)}\right]\right\}\, ,
\end{split}
\end{equation}
with $\lambda=\mbox{sign}(\nu)$, and therefore contains different powers of $\log\eta$ as well as some functions of $\xi^2$. However, when we add the corresponding asymptotics of $M^{(3)}_{1,\sigma_1\sigma_2\sigma_3}$ which is the leading term in the limiting case under discussion, any dependence on $\eta$ and $\xi$ is cancelled out and we obtain the high-energy asymptotics of photon-photon scattering in vacuum \cite{BKKF74}. Therefore, analogously to the case of photon-photon scattering in vacuum, the high-energy asymptotics of the amplitudes of photon splitting in a laser field is independent of the energy of the initial photon as well as of the parameter $\xi$. More precisely, these statements are valid for any value of $\xi$ which satisfies the inequality $\xi^2\ll\eta$.

\section{Possibility of  experimental observation of photon splitting in a laser field}
We now discuss the possibility of experimental observation of photon splitting in a laser field. It is natural to consider the principal scheme of the experiment which was used in the first successful observation of photon splitting in a Coulomb field \cite{splitexp}. In this experiment a tagged photon beam was used and the energy of each final photon was measured by a calorimeter. It was necessary to confirm two-photon events, and the distance $d$ between photons in the calorimeter was about one centimeter. Assuming a distance $l\approx 25\; \mbox{m}$ between the calorimeter and the interaction region between tagged photons and laser field, then the angles $\vartheta_{2}=\angle (\bm k_2, \bm k_1)$ and $\vartheta_{3}=\angle (\bm k_3, \bm k_1)$ have to be larger than $\vartheta_0=4\times 10^{-4}\;\text{rad}$. At $\omega_1\gg\omega_0$ the final photons are almost parallel to the initial one so that $\vartheta_{2,3}\ll 1$. It follows from  Eq. (\ref{theta23}) that in this limit
\begin{equation}
\vartheta_2=2\sqrt{\frac{\Omega_n\nu_3}{\omega_1\nu_2}}\, ,\quad \vartheta_3=2\sqrt{\frac{\Omega_n\nu_2}{\omega_1\nu_3}}\, .
\end{equation}
If, for instance, $\nu_2>\nu_3$, then for $n=1$
\begin{equation}
\omega_1 < \frac{4\nu_3\omega_0}{\nu_2\vartheta_0^2}=25\times 10^6\,\frac{\nu_3\omega_0}{\nu_2}\,.
\end{equation}
For the planned photon energy of the X-FEL at DESY \cite{XFEL}, with $\omega_0=1$ keV, the latter condition becomes $\omega_1<25\nu_3/\nu_2$~GeV and is not restrictive. Then, we can essentially reduce $l$ or $\omega_1$, or increase $\vartheta_0$. The planned intensity of the X-FEL is about $2\times 10^{15}\;\mbox{W/cm}^2$ corresponding to $E/E_c\approx 10^{-7}$ and $\xi\approx 5\times 10^{-5}$. In this case, the amplitudes of photon splitting coincide with those of photon-photon scattering. Therefore, it is convenient to choose $\omega_1=m^2/\omega_0=250$~MeV for which $\eta=1$. This value of $\eta$ corresponds to the threshold of electron-positron pair production and therefore the background initiated by this process will be strongly suppressed. Besides, at $\eta=1$ the cross section of photon-photon scattering is almost maximal and is given by $\sigma\sim 10^{-30}\;\mbox{cm}^2$ (see e. g. \cite{LL} on pg. 572). Now, from Eq. (\ref{flux}) we find $\dot W\approx 60\,\mbox{sec}^{-1}$. The X-FEL pulses will have a duration of about 100 fs and the interval between pulses will be about 93 ns. If we take, for example, a flux of $10^8$ tagged photons per second \cite{Tagged}, we obtain about two photon splitting events per hour. However, the very small laser beam-size makes this estimation essentially smaller. In consequence, the possibility of experimental observation of photon splitting in a laser field becomes problematic when using the X-FEL and the scheme of Ref. \cite{splitexp}.

We discuss now the possibility of using a very strong optical laser as that described in \cite{ELI}. In that case $\omega_0=1\,\mbox{eV}$, and the laser intensity will be of order of $10^{25}\;\mbox{W/cm}^2$ corresponding to $E/E_c\approx 4.6\times 10^{-3}$. For a  value of the angle $\vartheta_0=10^{-4}$ we obtain, e. g. for $\nu_3/\nu_2=1/4$, approximately $100\;\mbox{MeV}$ as the upper limit for $\omega_1$. For these parameters we find $\eta=4\times 10^{-4}$ and $\chi=0.9$. Therefore, we have to use Eq. (\ref{M3as}) but not the asymptotics (\ref{n1}) because high-order corrections in $\chi$ are important. Using the numerical results shown in Fig. \ref{fig:chi} we obtain $\dot W=4\times 10^8\, \mbox{sec}^{-1}$ at $\chi=0.9$. Though the value of $\dot W$ is very large, the duration of each pulse is $10$ fs and the laser repetition rate is only $1$ Hz \cite{ELI}. For a total flux of $10^8$ photons per second \cite{Tagged} and for an electron bunch revolution frequency equal to $1$ MHz in the accelerator, we obtain an average number of tagged photons per pulse equal to $100$. If all the photons pass through the laser pulse, we obtain again approximately two events per hour. However, the transverse beam size of a focused optical laser is even smaller than that of an X-FEL and the reduction due to the tagged and laser beams overlapping is larger.

Our negative conclusion about the possibility of observing photon splitting in a laser field when using the scheme discussed is mainly due to the short duration and small size of the laser beam as well as to the small flux of tagged photons. On the one hand, short and focused laser beams are closely coupled with large intensity of the laser pulses. On the other hand, the use of tagged photons, which essentially restricts the statistics, is needed to suppress the large background due to many other processes (double Compton scattering with the experimental equipment like collimators, bremsstrahlung accompanied by $e^+-e^-$ pair production and so on). Of course, we cannot exclude the existence of other schemes of the experiment that avoid the difficulties pointed out in this Section.

\section{Conclusions}
In conclusion, we have derived compact expressions for the amplitudes of the photon splitting process in a plane wave of arbitrary form, intensity and polarization (see Eqs. (\ref{S2final}) and (\ref{S3final})). In the simpler case of a monochromatic circularly polarized laser field, the amplitudes have the form of three-fold integrals (see Eqs. (\ref{M2circular}) and (\ref{M3circular})). Moreover, we analyzed in detail various asymptotics of the amplitudes in the case of a monochromatic wave with linear or circular polarization. We have demonstrated that the angular distribution of the probabilities corresponding to different numbers of absorbed photons is noticeably different; this permits us the distinction between these processes for the case where the laser photon energy is less than the accuracy of the final photon energy measurement. 

The photon splitting amplitudes depend on the two Lorentz-invariant parameters $\eta$ and $\chi$. For arbitrary values of these parameters, the amplitudes both in the case of a circularly and a linearly polarized laser field are not equal to zero for any (odd) number $n$ of absorbed laser photons. This is because the angle between the two final photons is not zero and the conservation of the projection $J_z$ of the total angular momentum (in the case of a circularly polarized laser field) does not imply any restriction. However, we have shown that in the limit of small $\eta$ the leading contribution to the amplitudes for a circularly polarized laser comes from the terms with $n=1$ and $n=3$, while for a linearly polarized laser field the leading contributions contain the terms with any odd $n$.

Finally, we have demonstrated that the scheme previously employed in the experiment of photon splitting in an atomic field together with new generation X-FEL sources and strong optical lasers is not suitable for the corresponding experiment in a laser field. Therefore, it is necessary to find another more appropriate experimental configuration to observe photon splitting in a laser field.

\section*{Acknowledgments}
We would like to thank K. Z.~Hatsagortsyan for valuable discussions. A. I. M. gratefully acknowledges the great hospitality and the financial support he has received during his visit at Max-Planck-Institute for Nuclear Physics. The work was supported in part by the RFBR under grants  0502-16079 and 06-02-04018.

\appendix*
\section{Coefficients for the helicity amplitudes}
Here we present the coefficients $B_{j,\sigma_1\sigma_2 \sigma_3}$ used in Eq. (\ref{M3circular}). We introduce the abbreviations
\begin{eqnarray}
&&x_j=\frac{s_j}{ S}\, ,\quad d_1=\frac{(\tau_1-\tau_2)f_2}{2\tau_1\tau_2}+\frac{(\tau_1+\tau_3)f_3}{2\tau_1\tau_3}\,,
\quad  d_2=\frac{(\tau_2-\tau_1)f_2}{2\tau_1\tau_2}+\frac{(\tau_1-\tau_3)f_3}{2\tau_1\tau_3}\,, \nonumber\\ 
&& d_3=\frac{(\tau_2+\tau_1)f_2}{2\tau_1\tau_2}-\frac{(\tau_1+\tau_3)f_3}{2\tau_1\tau_3}\,,  
\quad h=d_1+d_2+d_3=\frac{(\tau_2+\tau_1)f_2}{2\tau_1\tau_2}+\frac{(\tau_1-\tau_3)f_3}{2\tau_1\tau_3}\, ,\nonumber\\
&&g_1=\frac{h}{\eta S}-1-ih\,,\quad g_2=\frac{h}{\eta S}-1-2if_2+i\nu_2h\,,\quad g_3=\frac{h}{\eta S}-1-2if_3+i\nu_3h\,.
\end{eqnarray}
Then, the nonzero coefficients for the contribution  $M^{(3)}_{n,+++}$  are 
\begin{eqnarray}\label{+++}
B_{2,+++}&=&4\rho_n\,\xi^2x_1\left(g_2 g_3+\nu_2\nu_3d_1^2 \right)\, ,\nonumber\\
&&\nonumber\\
B_{1,+++}&=&4\xi^3\left[g_1^* g_2 g_3+\nu_2\nu_3 g_1^*d_1^2+(\nu_3g_2+\nu_2g_3)|h|^2-i\nu_2\nu_3h^*(h^2+d_1^2)\right]\nonumber\\
&&-4\rho_n^2\xi x_1\left(\frac{x_2 g_2}{\nu_3}+\frac{x_3 g_3}{\nu_2}+i d_1\right)-\frac{4i\xi}{ S}(g_2+g_3)\, ,\nonumber\\
&&\nonumber\\
B_{0,+++}&=&-4\rho_n\xi^2\Bigg[g_1^*\left(\frac{x_2g_2}{\nu_3}+\frac{x_3g_3}{\nu_2}\right)
+ig_1^*d_1  +h^*d_1+\left(\frac{\nu_2x_2}{\nu_3}+\frac{\nu_3x_3}{\nu_2}\right)|h|^2\Bigg] \nonumber\\
&&+\frac{\rho_n^3x_1(4x_2x_3-1)}{\nu_2\nu_3}+\frac{4i\rho_n}{ S}\left(\frac{x_2}{\nu_3}+\frac{x_3}{\nu_2}\right)\, ,\nonumber\\
&&\nonumber\\
B_{-1,+++}&=&\frac{\rho_n^2\xi}{\nu_2\nu_3}\left[(4x_2x_3-1)g_1^*+ih^*\right] \, ,
\end{eqnarray}
The nonzero coefficients for the contribution  $M^{(3)}_{n,++-}$  are 
\begin{eqnarray}\label{++-}
B_{1,++-}&=&\frac{\rho_n^2\xi}{\nu_3}[(4x_1x_2-1)g_2+i\nu_2 h]\, ,\nonumber\\
&&\nonumber\\
B_{0,++-}&=&4\rho_n\xi^2\left[g_2\left(\frac{x_2 g_1^*}{\nu_3}-x_1g_3^*\right)
-ig_2d_3^*  -\nu_2hd_3^*+\nu_2\left(\frac{x_2}{\nu_3}+x_1\nu_3\right)|h|^2\right]\nonumber\\ 
&&-\frac{\rho_n^3x_3(4x_1x_2-1)}{\nu_2\nu_3}+\frac{4i\rho_n}{ S}\left(x_1-\frac{x_2}{\nu_3}\right)\, ,\nonumber\\
&&\nonumber\\
B_{-1,++-}&=&4\xi^3\left[-g_1^*g_2g_3^*+\nu_2(\nu_3g_1^*-g_3^*)|h|^2+\nu_3g_2d_3^{*2}-i\nu_2\nu_3h(h^{*2}+d_3^{*2})\right] \nonumber\\
&&+\frac{4\rho_n^2\xi x_3}{\nu_2} \left(x_1 g_3^*-\frac{x_2}{\nu_3}g_1^*+id_3^*\right)
+\frac{4i\xi}{ S}(g_1^*+g_3^*)\, ,\nonumber\\
&&\nonumber\\
B_{-2,++-}&=&\frac{4\rho_n\xi^2x_3}{\nu_2}\left(g_1^*g_3^*-\nu_3d_3^{*2}\right)\, .
\end{eqnarray}
Finally, the nonzero coefficients for the contribution  $M^{(3)}_{n,+--}$  are 
\begin{eqnarray}\label{+--}
B_{0,+--}&=&\frac{4x_1x_2x_3\rho_n^3}{\nu_2\nu_3}\, ,\nonumber\\
&&\nonumber\\
B_{-1,+--}&=&\rho_n^2\xi\Bigg[\frac{4x_2x_3-1}{\nu_2\nu_3}g_1^*-\frac{4x_1x_2-1}{\nu_3}g_2^* -\frac{4x_1x_3-1}{\nu_2}g_3^* \nonumber\\
&&-2i(1-2x_1)d_1^*+\frac{2i(1-2x_3)}{\nu_2}d_3^*+\frac{2i(1-2x_2)}{\nu_3}d_2^*\Bigg] \,,\nonumber\\
&&\nonumber\\
B_{-2,+--}&=&4\rho_n\xi^2\Bigg[ x_1g_2^*g_3^*- g_1^*\left(\frac{x_2}{\nu_3}g_2^*+\frac{x_3}{\nu_2}g_3^*\right)+ig_1^*d_1^*+ig_2^*d_3^*
+ig_3^*d_2^*\nonumber\\  
&&+x_1\nu_2\nu_3d_1^{*2}+\frac{x_2\nu_2}{\nu_3}d_2^{*2}+\frac{x_3\nu_3}{\nu_2}d_3^{*2}+\nu_2d_1^*d_2^*
+\nu_3d_1^*d_3^* -d_2^*d_3^* \Bigg]\, ,\nonumber\\
&&\nonumber\\
B_{-3,+--}&=&4\xi^3( g_1^*g_2^*g_3^*+\nu_2\nu_3g_1^*d_1^{*2}-\nu_3g_2^*d_3^{*2}-\nu_2g_3^*d_2^{*2}
+2i\nu_2\nu_3d_1^*d_2^*d_3^* )\, .
\end{eqnarray}
The coefficients for the amplitude  $M^{(3)}_{n,+-+}$ can be obtained from  $M^{(3)}_{n,++-}$ by the replacement  $2\leftrightarrow 3$. The coefficients for the amplitudes  $M^{(3)}_{n,---}$, $M^{(3)}_{n,-+-}$ and $M^{(3)}_{n,-++}$ can be obtained with the help of the substitutions
\begin{eqnarray}
&&B_{j,-++}=-B_{-j,+--}^*( S\to - S)\, ,\quad B_{j,-+-}=-B_{-j,+-+}^*( S\to - S)\,,\nonumber\\
&& B_{j,---}=-B_{-j,+++}^*( S\to - S)\,,
\end{eqnarray}
where the replacement $S\to- S$ means that after complex conjugation, it is also necessary to change the sign of the terms containing $S$ in Eqs. (\ref{+++}), (\ref{++-}) and (\ref{+--}).

\clearpage

\clearpage

\clearpage

\clearpage

\begin{figure}
\begin{center}
\includegraphics[width=10cm]{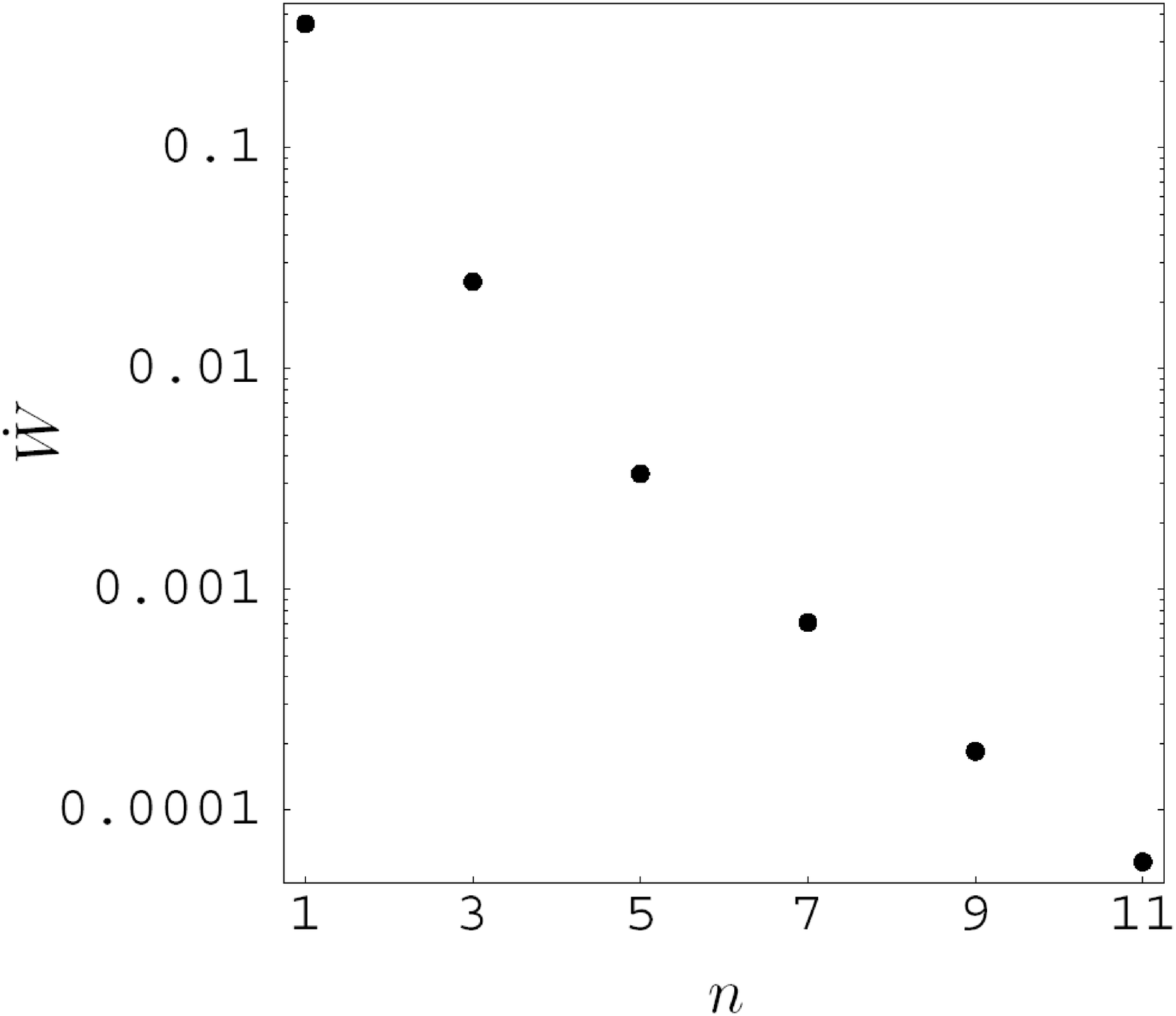}
\end{center}
\caption{The photon splitting rate $\dot W$ in a linearly polarized laser field in units $(\pi^2m^2\alpha^3/\omega_1)\times 10^{-4}$ as a function of the number $n$ of laser photons absorbed. The rate is calculated from the amplitudes in Eqs. (\ref{ML}) with $\chi=1$, is averaged over the polarization of the initial photon, summed up over the polarizations of the final ones and integrated over $\nu_2$ and $\phi_q$.}\label{fig:n}
\end{figure}

\end{document}